\documentclass[12pt, draftcls,onecolumn]{article}

\usepackage{enumitem}
\usepackage{rotating} 
\usepackage{amsmath}
\usepackage{algorithm}
\usepackage{algpseudocode}
\usepackage{amsmath}
\usepackage{etoolbox}
\usepackage{booktabs}
\usepackage{multirow}

\usepackage{url}
\usepackage{mathtools, cuted}
\usepackage{stfloats}
\usepackage{hyperref}
\usepackage{amsmath,amsfonts,amssymb}
\usepackage{bbm}

\usepackage{wrapfig}
\usepackage{draftwatermark}
\usepackage{algorithm}
\usepackage{algpseudocode}
\SetWatermarkText{}
\SetWatermarkScale{1}

\newtheorem{example}{Example}
\newtheorem{remark}{Remark}

\usepackage{bm}
\usepackage{mathptmx} 
\usepackage{amsmath}  
\usepackage{amssymb}  
\usepackage{amsfonts}
\usepackage{boxedminipage}
\usepackage{graphicx}
\usepackage{subcaption}  
\usepackage[T1]{fontenc}
\usepackage[english]{babel}
\usepackage{pst-sigsys}
\usepackage{pst-node,pst-circ}

\definecolor{dgray}{gray}{0.6}
\definecolor{lgray}{gray}{0.8}
         
\newcommand{\ignore}[1]{}

\def\bbbz{{\mathchoice {\hbox{$\sf\textstyle Z\kern-0.4em Z$}}
{\hbox{$\sf\textstyle Z\kern-0.4em Z$}}
{\hbox{$\sf\scriptstyle Z\kern-0.3em Z$}}
{\hbox{$\sf\scriptscriptstyle Z\kern-0.2em Z$}}}}

\newcommand{\ba}{\begin{array}}
\newcommand{\ea}{\end{array}}
\newcommand{\beq}{\begin{equation}}
\newcommand{\eeq}{\end{equation}}
\newcommand{\beqy}{\begin{eqnarray}}
\newcommand{\eeqy}{\end{eqnarray}}
\newcommand{\beqyn}{\begin{eqnarray*}}
\newcommand{\eeqyn}{\end{eqnarray*}}
\newcommand{\bi}{\begin{itemize}}
\newcommand{\ei}{\end{itemize}}
\newcommand{\bseq}{\begin{subeqnarray}}
\newcommand{\eseq}{\end{subeqnarray}}

\newcommand{\bex}{\begin{example}}
\newcommand{\eex}{\end{example}}
\newcommand{\bexer}{\begin{exercise}}
\newcommand{\eexer}{\end{exercise}}
\newcommand{\bmct}{\begin{fact}}
\newcommand{\efct}{\end{fact}}

\DeclareMathOperator*{\argmax}{argmax}

\patchcmd{\IEEEproofindentspace}{2\parindent}{-2pt}{}{}

\usepackage{authblk}

\date{}

\begin{document}

\title{
Surge sourcing via hybrid supply in a sharing economy: a resource-efficient, progressive and sustainable way to satisfy surge demand
}

\author[1]{Pouria Mohamadzadehoqaz}
\author[1]{Elena Dieckmann}
\author[1]{Anthony Quinn}
\author[1]{Robert Shorten}

\affil[1]{\small{Dyson School of Design Engineering, Imperial College London, South Kensington, UK}}

\maketitle

\vspace*{-.4cm} 

\begin{abstract}

We propose a surge sourcing approach to address occasional synchronous high demand (surge demand) in sharing economy systems, providing a socio-economically progressive alternative to surge pricing. Instead of suppressing demand among disadvantaged consumers, our scheme increases supply by involving privileged consumer-providers (prosumers) who under-utilize their resources. This hybrid supply approach maintains high quality-of-service (QoS) for both consumers and prosumers in both normal and surge demand situations without surge pricing. To ensure prosumer QoS, we reserve a small portion of the primary supply to meet their needs if their resources become unavailable during surge periods. As the probability of such events is low compared to that of the surge demand itself, the reserved resources required are minimal. The resulting scheme is resource-efficient, socially progressive, and sustainable, exploiting under-used resources. We illustrate our scheme through two applications: high-range car sharing for owners of small EVs, and shared charging points for EV drivers.

\end{abstract}

\noindent\textbf{Keywords:} Sharing economy,  hybrid Supply system, Surge sourcing, Prosumer, Quality-of-Service (QoS), Probability modelling

\section{Introduction}
Currently, there is much interest in finding economic models that disassociate economic growth from resource consumption. In this context, both the Sharing Economy \cite{Heras2020Analysis, Schneider2019Sustainability} and the Circular Economy \cite{Merli2017How,Androniceanu2021Circular} are seen as progressive economic models that move toward this goal. While the Circular Economy seeks to achieve a complete decoupling of growth and consumption, Sharing Economy systems are more modest in their objectives, being primarily concerned with resource optimisation by rethinking ownership and access models. {\em Sharing Economy} systems refer to systems that seek to develop new ownership models with the objective of replacing {\em outright ownership} with an {\em access-based} model. These models, while not without problems, are viewed as being more environmentally friendly, as they reduce the demand for resources, socially progressive from the point of view that access-based ownership models alleviate {\em access poverty} by providing consumers with temporary access to expensive goods and services, without the need for capital investment. Other benefits include the reduction of demand on raw materials through efficiency increase in resource usage .\newline

The objective of this paper is to develop an analytic tool that can be utilized to design sharing economy systems. In particular, we are interested in an emerging class of sharing economy systems in which multiple shared services serving distinct consumer groups are coupled via a single shared resource. In such systems, the {\em quality-of-service} (QoS) delivered to group $A$ may, and often does, affect the QoS delivered to group $B$, and vice versa.\newline

Examples of situations in which coupled sharing economy systems can be identified in a range of different contexts. An emerging problem in the automotive world is the issue of EV obesity \cite{wieberneit2024optimal}. This refers to a situation whereby the weight of electric vehicles is increasing due to consumer demand for increased battery capacity. This development is problematic as it leads to increased non-tailpipe emissions (tyres, brakes, road dust, road markings) \cite{Harrison2021Non-exhaust, Timmers2016Non-exhaust, Piscitello2021Non-exhaust, Beddows2021PM10}, decreased vehicle efficiency, and makes these vehicles expensive for consumers. Paris has introduced higher parking fees for larger vehicles\footnote{\href{https://urban-mobility-observatory.transport.ec.europa.eu/news-events/news/paris-introduces-triple-parking-fees-suvs-2024-02-12_en}{urban-mobility-observatory.transport.ec.europa.eu}}. In practice, this problem is difficult to solve due to journeys in which electric car owners require access to vehicles with longer ranges (and thus larger and heavier batteries). One solution is to provide consumers flexibility via a hybrid ownership model; that is, a community of $N$ consumers, when purchasing small vehicles, automatically gain access to a fleet of $M$ vehicles with larger batteries. As longer journeys for city dwellers are rarer than short ones, it is reasonable to expect that a small number of such shared vehicles would be sufficient to serve the needs of the larger community \cite{21}. While such a system may be viewed as a form of battery swapping, there are commonly periods of synchronized demand during the provision of shared vehicles that may not be sufficient to serve the needs of the community of small-vehicle owners. Synchronized demand may be served in three ways; (i) via {\em dynamic, or surge, pricing} \cite{2}; (ii) via a policy of {\em fairly allocating resources}; and (iii) via sourcing, over short time scales, additional larger vehicles. Surge pricing is currently the method of choice for managing excess demand in such situations. However, this is also a very socially regressive form of demand management as it curtails access to a resource based on the ability of an individual to pay. An alternative is the use of fair access policies that give access to a resource based on some community-agreed fairness criteria. However, over repeated periods of excess demand, such policies may achieve group fairness by delivering a very low QoS to all consumers over time. An alternative to both of the above is to source excess idle supply through consumers, "prosumers" ($T$), during periods of excess demand. We refer to this approach as {\em surge sourcing}. A significant barrier to surge sourcing is the {\em angst} that vehicles sourced from a population $T$ will not be returned on time for use by their owners, limiting their ability to execute desired functions. In our system design, we develop a reserve mechanism (Q) in a principled manner to reassure the population $T$ that they will have access to their resource in the case when their own asset is not returned promptly or a spontaneous need for usage arises when their item is in the shared pool (M). The pool of shared items can be described as $M-Q$ to deliver a QoS to a group of users $N$, in the case of the example above, the community of small vehicle owners. The system we propose has two shared communities, of dimension $N$ and $T$, coupled via $M-Q$ and $Q$ shared objects, where the QoS to $N$ and $M$ affect each other in an adversarial manner.  \newline

Our contribution is an analytic framework for the design of hybrid supply systems. This framework represents an extension of the state-of-the-art in that several sharing economy systems are coupled via a single shared resource. We use a shared resource to deliver a specified QoS to a population and also to reduce risk to prosumers, who provide additional resources during periods of surge demand, offering an alternative to surge pricing. Our system has economic benefits for consumers due to a maximized QoS, to $T$, the population of prosumers who can make gains by loaning their assets temporarily, for the operator of $M$ due to flexibility in dimensioning infrastructure through access to additional assets of $T$. 

\subsection{Related work}

Several communities have worked on analytics that have been useful for the design of sharing economy systems. In particular, recent work on consensus protocols \cite{saber2003consensus}, on distributed optimization \cite{alam2020distributed, nedic2014distributed}, and on the design of distributed ledgers (blockchain), are three examples of domains areas that have aided the design of sharing economy systems. In terms of specific work, the papers most related to the work presented here are \cite{18} and \cite{21}. The study \cite{21} addresses electric vehicle (EV) range anxiety by proposing an on-demand vehicle access model that supplements EV ownership with access to shared internal combustion engine (ICE) vehicles, especially for long trips exceeding the EV range. Using queueing theory, the researchers developed a fleet model that ensures high QoS at minimal additional cost, making EV adoption more attractive, particularly for city dwellers with predictable, short-range travel needs. However, a potential issue is the study's suggestion of dynamic pricing to manage surge demand, which could be perceived as unfair to users, particularly those who rely on the service during peak times. A related paper \cite{18} explores campus parking systems with QoS guarantees, focusing on optimizing the use of parking spaces through shared arrangements between campus and nearby residential areas. Two design challenges are addressed: ensuring daytime contingency spaces for residents and improving parking allocation efficiency. QoS metrics are also used in  \cite{23} on Electric Vehicle (EV) to explore how unused battery capacity in EVs could mitigate the risks associated with intermittent renewable energy sources like wind and solar. Intermittency introduces uncertainty in balancing supply and demand on the grid, which can hinder renewable energy investment and growth. Instead of focusing on pricing, the research examines the scale of an EV fleet required to provide reliable backup storage. By using QoS metrics, the findings indicate that only a small number of participating EVs is needed to mitigate production risks, with minimal impact on these vehicles. Sharing economy analytics are also developed in \cite{22}, which presents a Smart Plug adapter and a Sharing System. The Smart Plug adapter is a hardware device that enables a single charge point to serve two vehicles, allowing private owners to rent their chargers to the public with integrated booking and payment features. The Sharing System employs a QoS-based approach to maintain reliable service, incorporating a "Q reserve" mechanism to address potential issues, such as charge point owners not vacating as scheduled. This reserve system ensures backup charge points are available in high-demand areas, supporting consistent access and alleviating charge point anxiety for EV users. The study \cite{Blockchain-based_charging_pile} proposes an energy blockchain-based secure sharing scheme for private charging pile networks to address operational challenges, enhancing EV user utility and renewable energy efficiency through simulations and real-world implementation. The study \cite{distribution_charging_pile} addresses the challenge of fair benefit distribution in private charging pile sharing by introducing an improved Shapley value model. This model considers key factors such as risk, input, and user-assessed service quality to ensure equitable sharing of benefits, helping alleviate infrastructure shortages in large cities. The study \cite{27} proposes a secure blockchain-based scheme for private charging networks to enhance user quality-of-experience (QoE) and protect against security threats.\\

Work on surge pricing can be found in several communities. For example, the study \cite{Two-sided-competition} analyzes competition between two-sided platforms in the sharing economy, focusing on the impacts of self-scheduled supply and wage schemes on platform competition. They model the interactions between workers (who provide assets) and consumers (who need assets) under various wage schemes: fixed commission rate, dynamic commission rate, and fixed wage. When demand-side competition exceeds supply-side competition, platforms benefit most from a fixed-wage scheme, though this is less favourable for consumers and workers. Conversely, intense supply-side competition favours platforms using a fixed-commission scheme when supply competition is very high or a dynamic commission scheme when supply competition is moderate. A somewhat related paper, \cite{Intermediated_surge_pricing}, investigates why platforms like Uber and Lyft apply flexible "surge pricing" but consistently take a constant percentage of each fare, even during demand surges. This intermediary approach sets prices for both buyers (riders) and sellers (drivers), keeping a fixed percentage as their fee. When the intermediary maintains a constant fee percentage amid demand shifts, surge pricing becomes more pronounced on one side of the market while being reduced on the other. The analysis employs the Cournot competition model, where firms independently and simultaneously decide on quantities to produce, to explore the effects of this rigidity. The study \cite{scarcity} explores scarcity—similar to surge demand in other research—in peer-to-peer sharing platforms using a game-theoretic model to examine how customer evaluations can support stable provider-user matching. It proposes that a user-proposing deferred acceptance algorithm on ride-sharing platforms with high user-to-provider ratios can effectively reduce congestion and ensure stable matches. By incorporating a customer satisfaction rating system, the model modifies the traditional batch matching approach, allowing highly-rated riders to be matched more quickly than lower-rated ones. However, a potential issue arises with the accuracy of ratings, as low ratings may not always be fair or reflective of actual behaviour, which could introduce bias and lead to unfair treatment of certain users.\\

\subsection{Contributions and layout of the paper }

This paper goes beyond the authors' related work  \cite{18} by proposing a reserve to alleviate prosumer angst. Specifically, our main contributions, here,  are as follows:
\begin{enumerate}
\item To the best of our knowledge, this is the first paper in which  a sharing economy system has been designed to include a primary sharing system, a reserve to alleviate angst, and a prosumer pool.
\item Several mathematical formulations of the complete system are provided. In the first of these, a mathematical optimization to dimension the overall system is presented. In the second,  the number of shared items is pre-specified, and algorithms to find the best possible split of the shared resource are presented. In particular, privacy-preserving algorithms based on the AIMD algorithm are provided. These yield an optimal allocation without the need to share sensitive information. 
\item Two significant use cases for our algorithms are presented. The first proposes a solution to the EV obesity problem. The second, based on real data from the company, Cirrantic \footnote{\href{https://cirrantic.com/}{cirrantic.com}}, develops an EV charge point sharing system.   
\end{enumerate}

The structure of the paper is as follows. In Section~\ref{sec:problem}, we introduce our proposed hybrid supply scheme and define two design problems associated with hybrid supply systems. Section~\ref{sec:QoS} formulates the QoS metrics mathematically, highlighting their significance in addressing the challenges outlined in Section~\ref{sec:problem}. Section~\ref{sec:opt} presents our methodologies for solving the design problems, detailing the optimization approaches employed. In Section~\ref{sec:res}, we apply the hybrid supply scheme to two use cases---car sharing and charge point sharing---and discuss the results. Finally, Section~\ref{sec:con} summarizes the key findings of this research. \newline

We adopt the following notational conventions. Specific populations involved in the hybrid supply scheme are usually denoted by $\mathcal N$ and $\mathcal T$, while cardinalities of populations and pools are denoted by $N$, $M$, {\em etc.}, which are all assumed to be finite. Probability models are all discrete, and are denoted by their first three letters, specifying their shaping parameters,  {\em e.g.}\ ${\mathcal B}er(p)$ for Bernoulli with probability $p\in(0,1)$.  If a random variable is independently and identically distributed according to a specified probability model, we annotate this fact by $\stackrel{iid}{\sim}$.

\section{Problem Statement}
\label{sec:problem}
We consider a community of $N$ consumers who may require access to a pool of shared items.

In this paper, we propose a methodology for designing a sharing scheme that satisfies the following requirements: it encourages the sharing of expensive and resource-intensive assets that are underutilized; it alleviates unfair and socio-economically regressive surge pricing; it

    provides a high level of satisfaction 
    to all stakeholders.\newline

Specifically, our {\em hybrid supply scheme} empowers a community of (primary) consumers with $N$ members to access a pool of $M$ shared items, where $M\ll N$. During normal periods, which we call {\em non-surge} demand (ns), this community accesses only the pool of shared items. In some circumstances, such as when demand is synchronized, which we call {\em surge demand} (s), consumers may access a supplementary pool of shared items. For example, these supplementary items may be provided, in a peer-to-peer manner, by individuals who own but under-utilize the items. The risk for these individuals during surge periods is that they may not be able to access their items when required (for example, outside of periods specified by a contractual agreement) due to the {\em bad behaviour} (b); {\em i.e.}\ non-compliance of the primary consumers. 
Therefore, referring to Figure 1, we have a population $\mathcal{T}$ of prosumers, which makes available an additional $T$ items to be used by population $\mathcal{N}$ (of $N$ consumers). In order to guarantee a satisfactory experience for these prosumers,  $Q$  of the $M$ items in the shared pool ($Q<M$) are reserved for the prosumers $\mathcal{T}$, to compensate for unexpected or `bad' behaviour on the part of any of some consumers in $\mathcal{N}$. 
\newline 

{\bf Example :} One use-case that embodies the structure in Figure 1 considers the sharing of high range electric vehicles (EVs) in an effort to reduce tyre emissions \cite{liu2022impact}. In this case, $\mathcal{N}$ is a population of owners of lightweight ({\em i.e.}\ with a small battery and therefore short range) EVs; in this case, $M$ is a pool of shared heavy (long range) EVs; and $\mathcal{T}$ is a population of long-range EV owners who choose, on occasion, to make their vehicles available to the shared pool for remuneration ({\em i.e.}\ an {\em Uber}-like usage model). For instance,  Tesla is currently instrumenting their vehicles and actively promoting a business model in which owners of their EVs are enabled to make them available to a peer-to-peer market \footnote{\href{https://www.reuters.com/technology/tesla-robotaxis-by-june-musk-turns-texas-hands-off-regulation-2025-02-10/}{reuters.com}}.\newline

\begin{figure}
    \centering
    \includegraphics[width=0.7\linewidth]{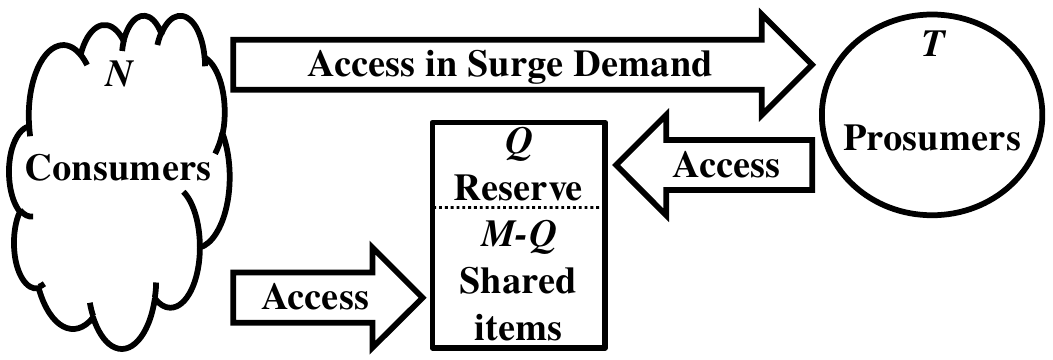}
    \caption{Our proposed hybrid supply scheme, in which  \(N\) consumers have regular access to a pool of \(M\) shared items during normal situations (non-surge demand periods) and gain access to \(T\) items from Prosumers during surge demand periods. Additionally, \(Q\) reserve items are available to support the \(T\) Prosumers during such periods.}
    \label{fig:Hybrid}
\end{figure}

\begin{remark}

It is important to note that the sourcing of additional supply from $T$ prosumers during periods of high asset demand---{\em i.e.}\ surge sourcing---is an alternative to surge pricing as a means of regulating the supply-demand balance. In our model (Figure~\ref{fig:Hybrid}), the $Q$ reserved items provide peace-of-mind for each prosumer in case the item that they made available to the shared pool---and which was used by a primary consumer---is not returned as planned.\end{remark}

To summarize, there are three demand scenarios associated with our scheme.

\begin{itemize}
    \item[] \textcolor{black}{Scenario $i=ns$:} Access of consumers with population N to $M$ items of the shared pool during non-surge demand periods.
    \item[] \textcolor{black}{Scenario $i=s$: } Access of consumers to $M-Q+T$ items, during surge demand periods.
    \item[] \textcolor{black}{Scenario $i=bs$: } Access of prosumers to $Q$ items, during surge demand periods. 
\end{itemize}

For each of these three scenarios,  we can define a measure of customer satisfaction that we call {\em quality-of-service} ($QoS$). This is a measure of how probable it is that a consumer/prosumer is able to access a shared resource when required.  Let $QoS_{ns}$, $QoS_s$, and $QoS_b$ be the appropriate QoS values associated with each of the three scenarios above as we shall now describe.  

\section{Formulation of QoS metrics for consumers and prosumers}
\label{sec:QoS}

We now formulate a QoS criterion for each of the three scenarios defined above ({\em i.e.}\ non-surge (ns), surge (s) and bad (b) behaviours, respectively). We require that all three criteria be satisfied by the optimization ({\em i.e.}\ the system design) in Section~\ref{sec:opt}. We make the following assumptions:
\begin{itemize}
    \item $QoS_i$ is to be expressed in probability, $i\in\{ns,\;s,\;b\}$;
    \item The number of (primary) consumers, $N$, is fixed.
    
    \item On any particular day, the request state, $R_{\cdot, i} \in \{1,0\}$, of the consumers and prosumers---encoding, respectively, whether they do or do not request an item from the scheme that day---are mutually independent and identically 
    distributed~\footnote{The iid assumption reflects the simplifying case in which---in any scenario---each consumer or prosumer has the same probability of requesting an item on a particular day, and is not influenced by any other.} Bernoulli trials, conditioned on $i\in\{ns,\;s,\;b\}$:
   \[
   R_{\cdot, i} \stackrel{iid}{\sim} {\mathcal B}er(p_i).
   \] 
    Here,  $p_i$ is the estimated probability that a request is made by a specific consumer or prosumer that day, conditioned on scenario $i\in\{ns,\;s,\;b\}$. In particular, $p_b$ is the probability that a prosumer will have to request an item from the reserve pool, $Q$ (Figure~\ref{fig:Hybrid}), owing to non-compliant ({\em i.e.}\ `bad') behaviour on the part of a consumer (such as the late return of an item to a prosumer), necessitating a request to $Q$. 
\end{itemize}
Under these assumptions, the total number of requests, $X_i \in \{0,1,\ldots\}$, conditioned on behaviour $i$, is binomial:
\begin{equation}
 X_i \sim {\mathcal B}in(n_i,p_i), \;\;i\in\{ns,\;s,\;b\}.
 \label{eq:bin}  
\end{equation}

The binomial parameters---for each of the three scenarios described above---are specified in Table~\ref{tab:QoS}

\begin{table}[h]
    \centering
    \scriptsize
    \caption{The quality-of-service parameters \eqref{eq:QoS_general_bin}, and their specified lower bound in each scenario.}
    \label{tab:QoS}
    \begin{tabular}{|c|c|c|c|c|}
        \hline
        \textbf{$i$ (Scenario indicator)} & \textbf{$n_i$ (Number in population)} & \textbf{$p_i$ (Probability of request)} &\textbf{ $A_i$ (Number of available items)} &\textbf{ $QoS_{i,d}$ (Lower bound)} \\
        \hline
        $ns$ (non-surge periods) & $n_{ns}=N$ & $p_{ns}$ & $A_{ns}=M$ & $QoS_{ns,d}$\\
        \hline
        $s$ (surge situations) & $n_{s}=N$ & $p_{s}$ & $A_{s}=M - Q + T$ & $QoS_{s,d}$\\
        \hline
        $b$ (bad behaviour) & $n_{b}=T$ & $p_{b}$ & $A_{b}=Q$ & $QoS_{b,d}$\\
        \hline
    \end{tabular}
\end{table}

An appropriate probability-based QoS is then specified as the probability that {\em all\/} the requests in each possible scenario, $i$, are satisfied, {\em i.e.}\ that $X_i \leq A_i$, where $A_i\in \{0,1,\ldots\}$ is the available number of items in scenario $i$ (Table~\ref{tab:QoS}). From (\ref{eq:bin}), these are computed as the binomial cumulative distribution function (cdf) evaluated at the respective $A_i$:
\begin{eqnarray}
QoS_i \;\;\equiv \;\; QoS_i(A_i; n_i, p_i) &\equiv&  \Pr[X_i \leq A_i | n_i, p_i, i] \nonumber\\
&&\nonumber\\
&\equiv& \left\{ 
\begin{array}{ll}
\displaystyle{\sum_{k=0}^{A_i} {n_i \choose k} p_i^k (1 - p_i)^{n_i - k}}, &A_i\leq n_i, \\
&\\
1, &A_i > n_i.
\end{array}
\right.
\label{eq:QoS_general_bin}
\end{eqnarray}

 In Section~\ref{sec:opt}, we will impose lower bounds on $QoS_i$, $\forall i$, to ensure that the system is designed to ensure minimal QoS in each scenario. In Section~\ref{sec:res}, we considered various values for $N$, including large values such as $N=50,000$. While a Poisson distribution is often more common and convenient for larger values of $N$, we choose to stick with the binomial distribution for the sake of consistency.

\begin{figure}[t]
    \centering
    \begin{subfigure}{0.49\textwidth}
        \centering
        \includegraphics[width=1.1\linewidth]{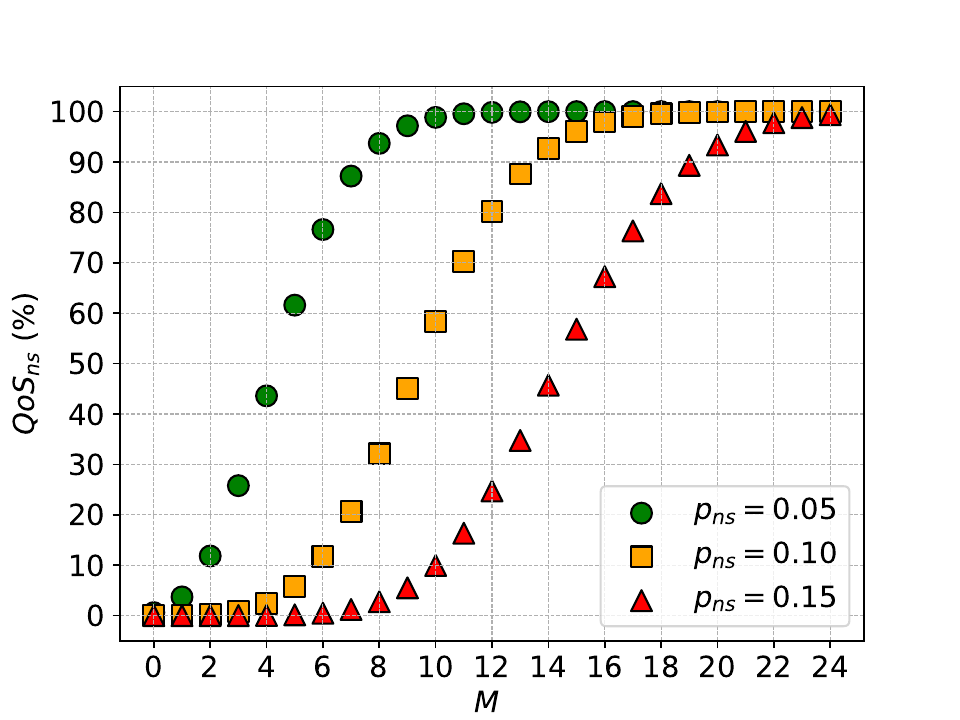} 
        \caption{$N=100$}
        \label{fig:QoS_ns(N_fixed)}
    \end{subfigure}
    \hfill
    \begin{subfigure}{0.49\textwidth}
        \centering
        \includegraphics[width=1.1\linewidth]{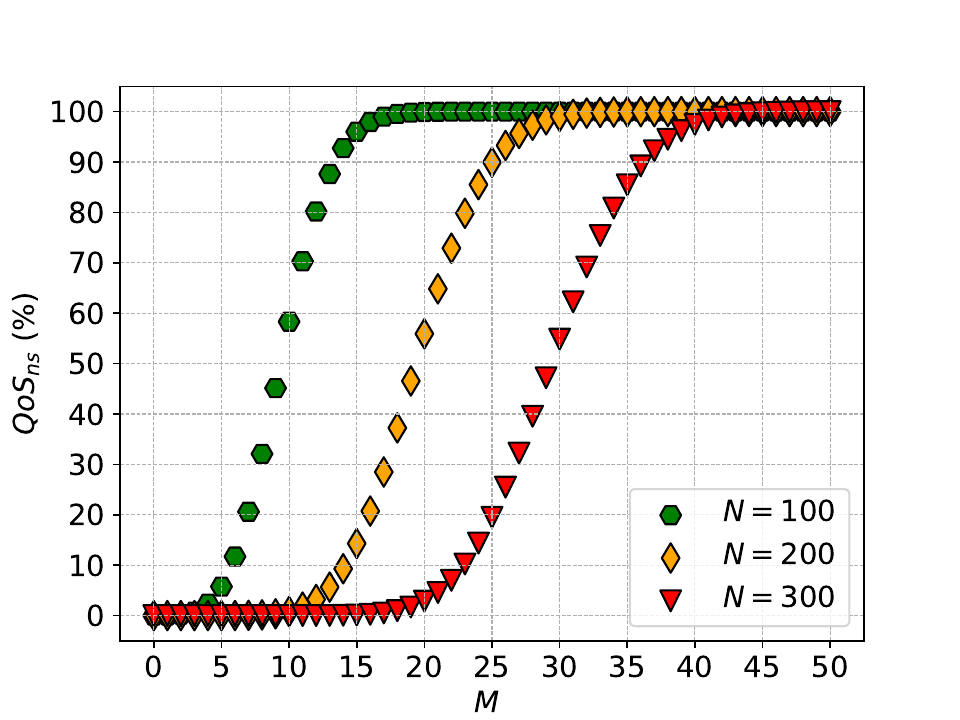}
        \caption{$p_{ns}=0.10$}
        \label{fig:QoS_ns(p_ns_fixed)}
    \end{subfigure}
    \caption{\small{$QoS_{ns}$ (Quality-of-service for consumers in a non-surge demand scenario) as a function of the number of items in the shared pool, \(M\). (a) Three different request probabilities, \(p_{ns}\), for fixed consumer population, $N=100$. (b) The effect of increasing $N$ for a fixed  $p_{ns}=0.10$.} 
    \label{fig:QoS_ns}}
\end{figure}

\begin{figure}[t]
    \centering
    \includegraphics[width=1\linewidth]{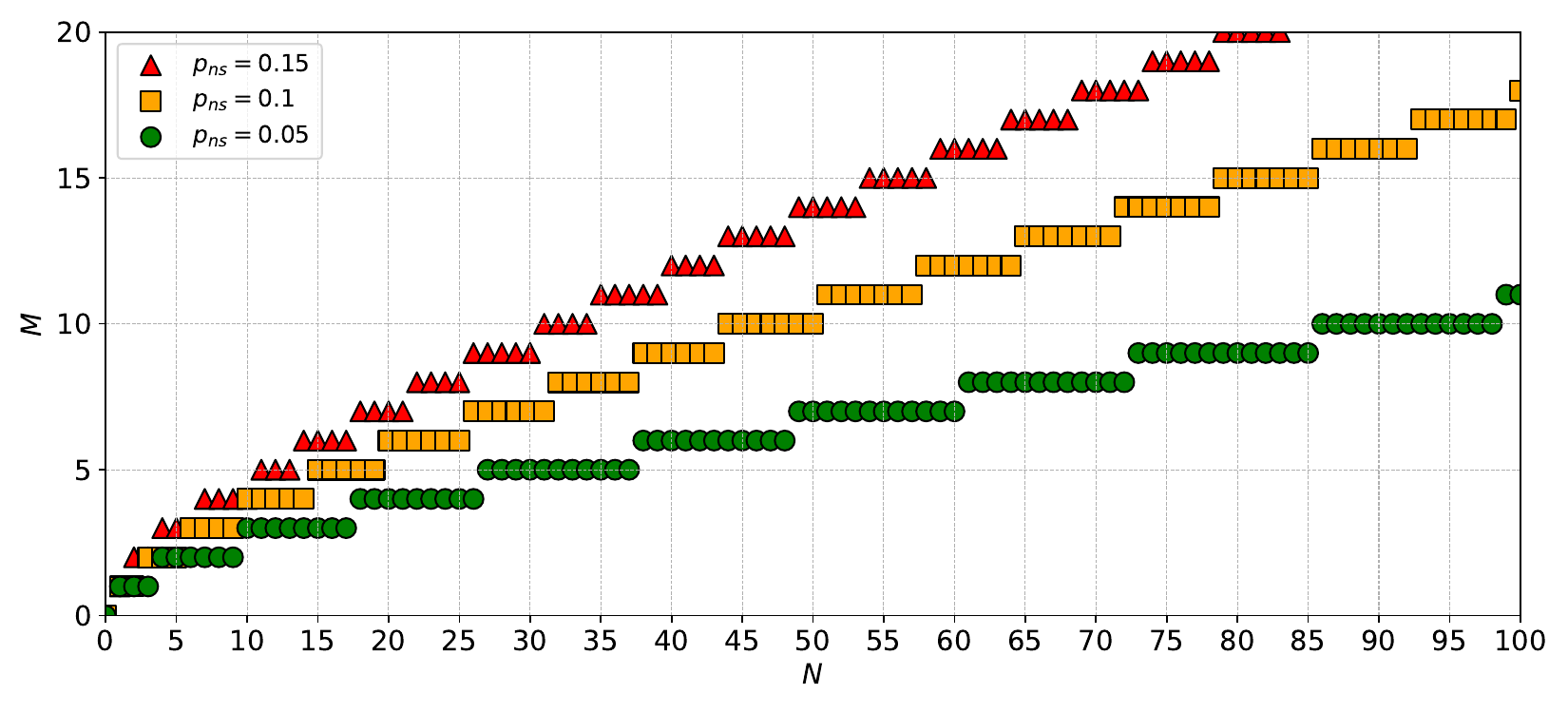}
    \caption{\small{$M$ (number of items in the shared pool) vs consumer population, $N$, for three request probabilities, $p_{ns}$, in a non-surge demand scenario. The figure displays the smallest $M$ for which  $QoS_{ns} \geq 99$\% in each case.}}
    \label{fig:QoS_ns(QoS_fixed)}
\end{figure}

\begin{figure}[htbp]
    \centering
    
    \begin{subfigure}{0.49\textwidth} 
        \centering
        \includegraphics[width=1\linewidth]{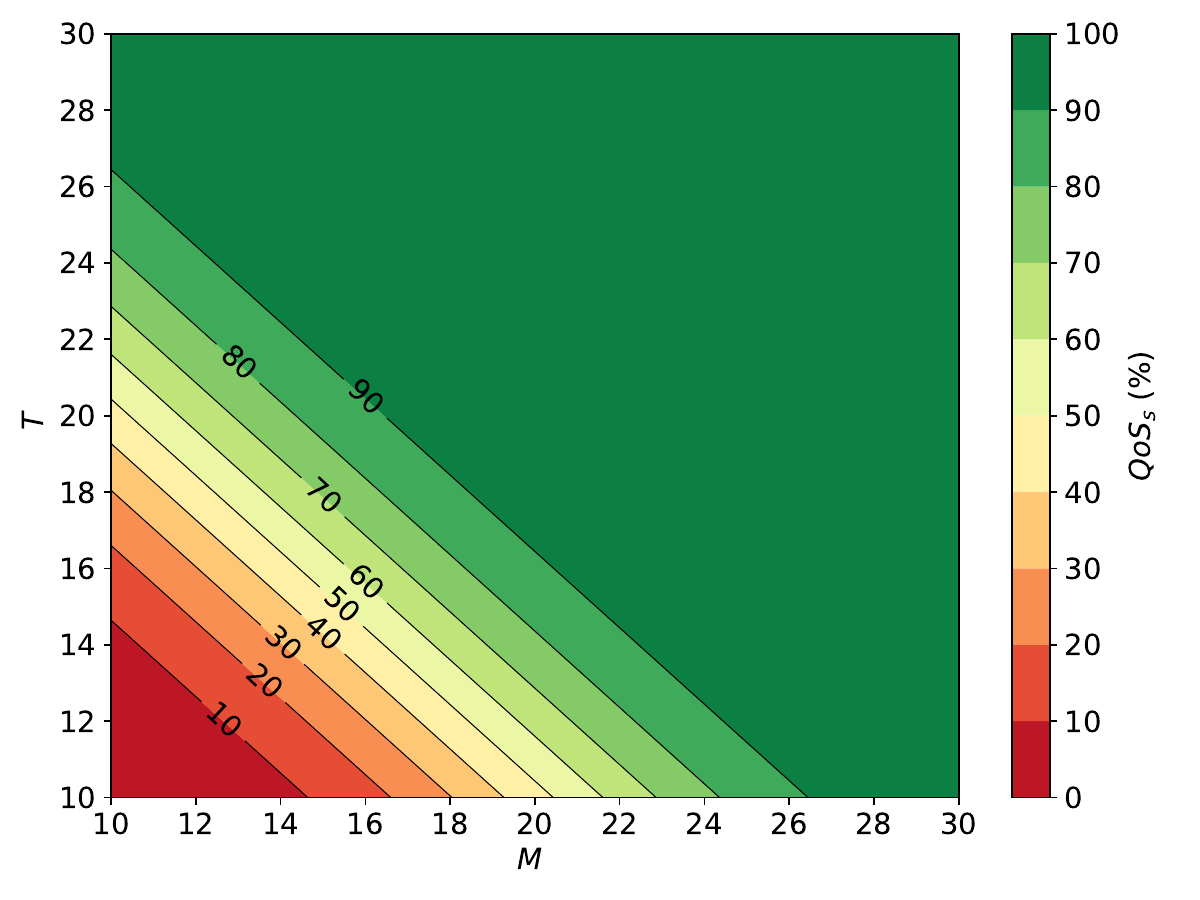} 
        \caption{$p_s=0.3, N=100, Q=1$}
        \label{fig:Top-Left}
    \end{subfigure}
    \hfill
    \begin{subfigure}{0.49\textwidth} 
        \centering
        \includegraphics[width=1\linewidth]{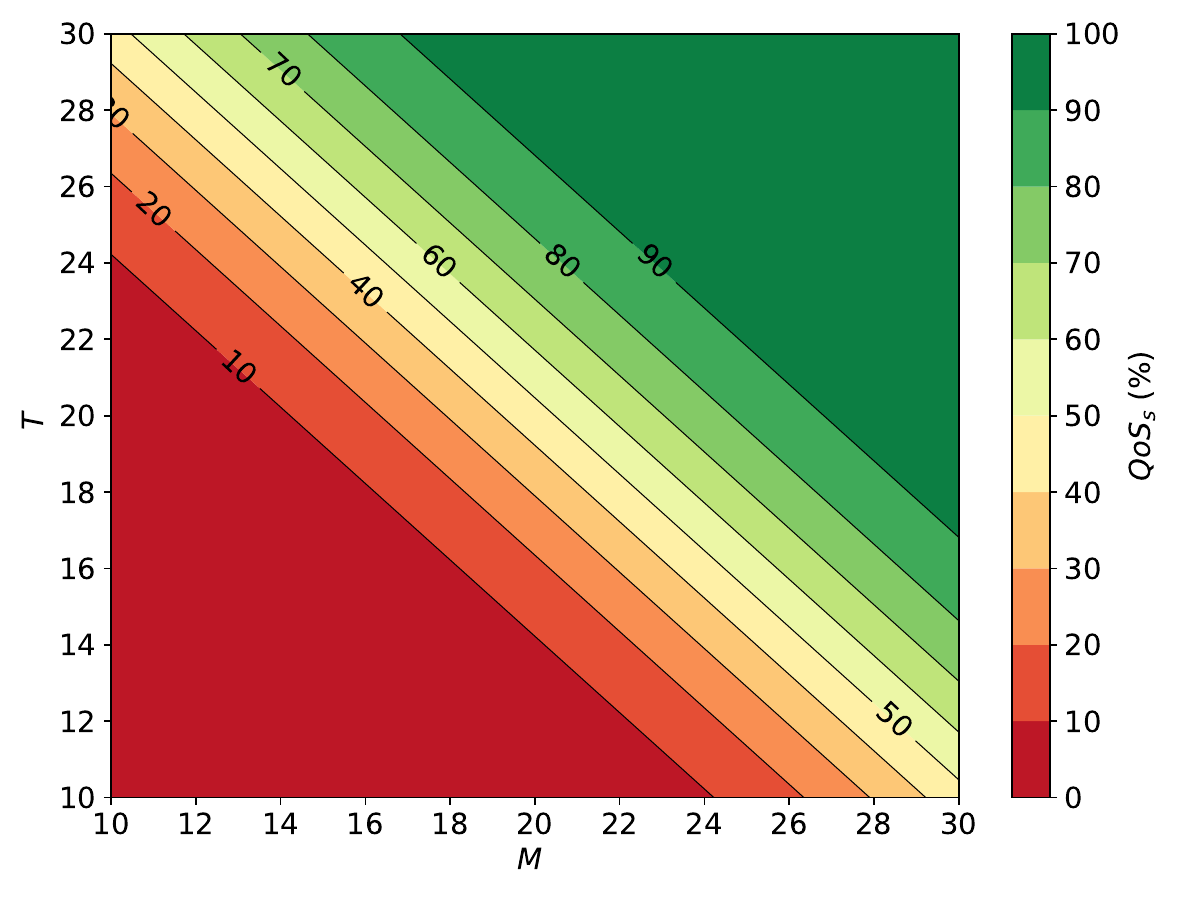} 
        \caption{$p_s=0.4, N=100, Q=1$}
        \label{fig:Top-Right}
    \end{subfigure}

    \vspace{1cm}
    
    \begin{subfigure}{0.49\textwidth} 
        \centering
        \includegraphics[width=1\linewidth]{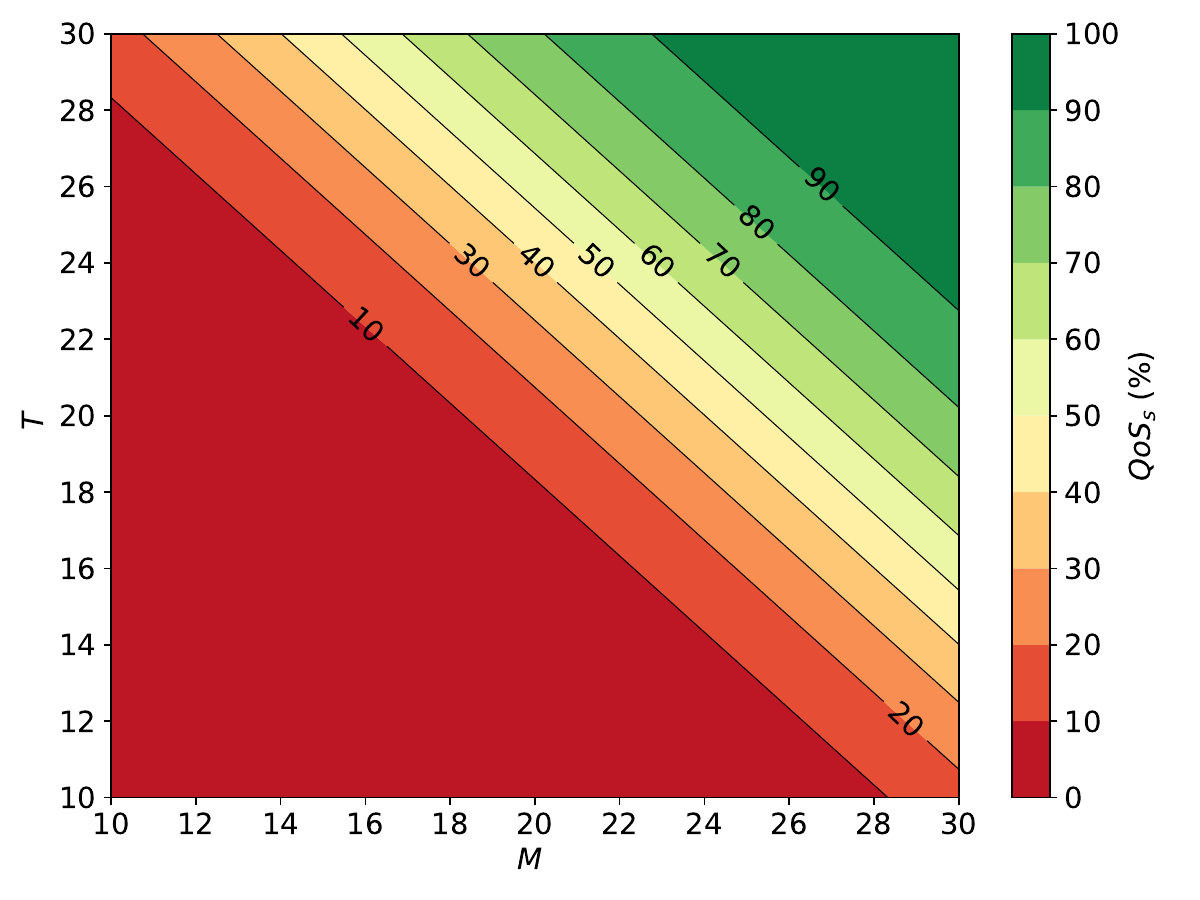} 
        \caption{$p_s=0.3, N=150, Q=1$}
        \label{fig:Bottom-Left}
    \end{subfigure}
    \hfill
    \begin{subfigure}{0.49\textwidth} 
        \centering
        \includegraphics[width=1\linewidth]{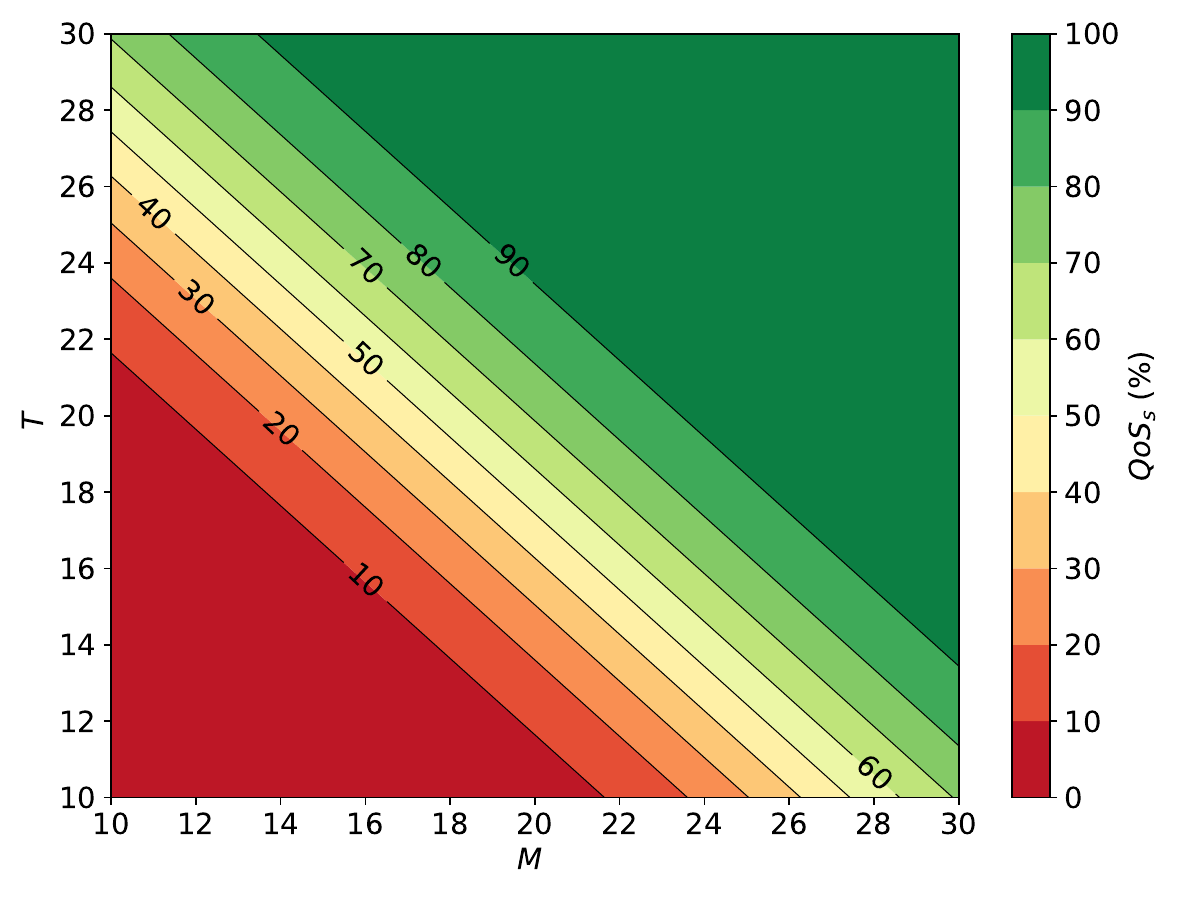} 
        \caption{$p_s=0.3, N=100, Q=8$}
        \label{fig:Bottom-Right}
    \end{subfigure}
    
    \caption{ \small{$QoS_s$ (quality-of-service for consumers during a surge demand scenario) as a function of $M$ (number of items in the shared pool) and $T$ (number of items from prosumers) for four different settings of the remaining QoS parameters (Table~\ref{tab:QoS}).}}
    \label{fig:QoS_s}
\end{figure}

\begin{figure}[htbp]
    \centering
    
    \begin{subfigure}{1\textwidth}
        \centering
        \includegraphics[width=1.1\linewidth]{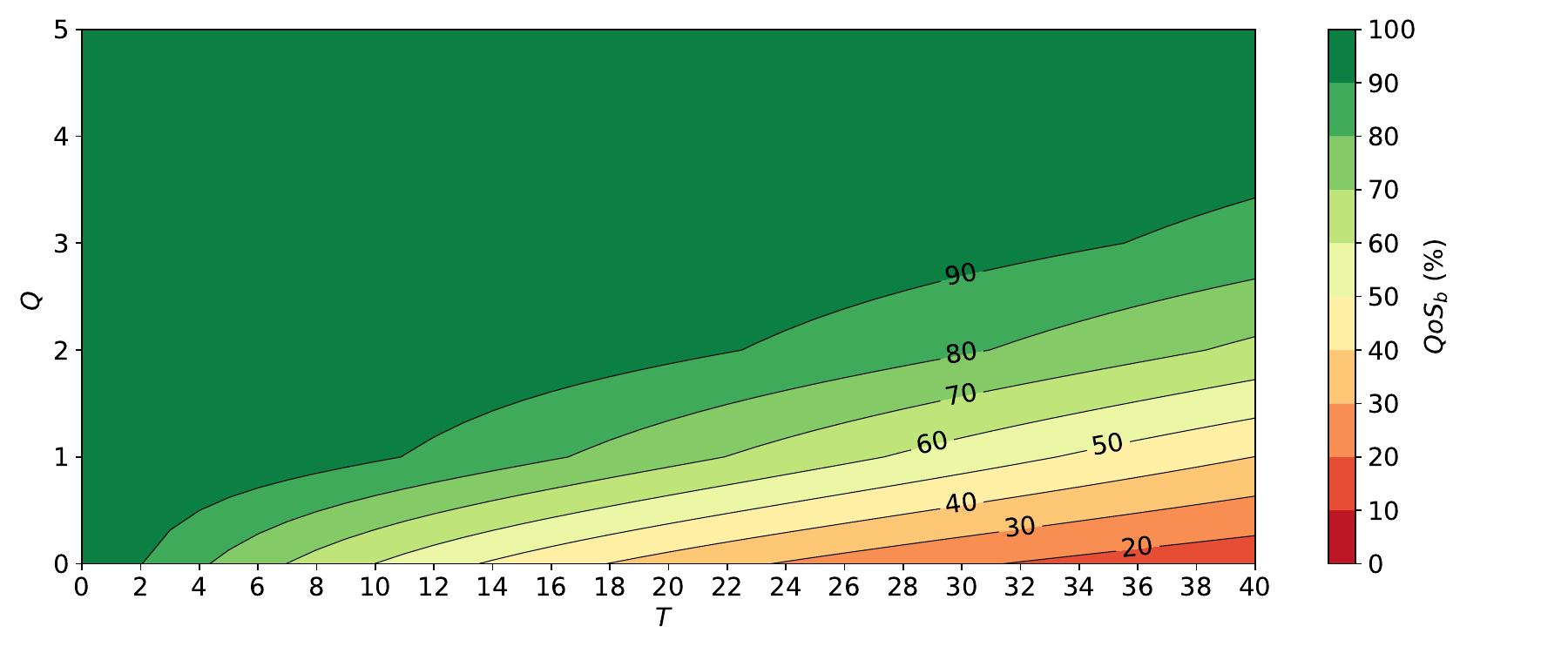} 
        \caption{$p_b=0.05$}
        \label{fig:QoS_b_0.05p_b}
    \end{subfigure}
    \hfill
    \vspace{1 cm}
    
    \begin{subfigure}{1\textwidth}
        \centering
        \includegraphics[width=1.1\linewidth]{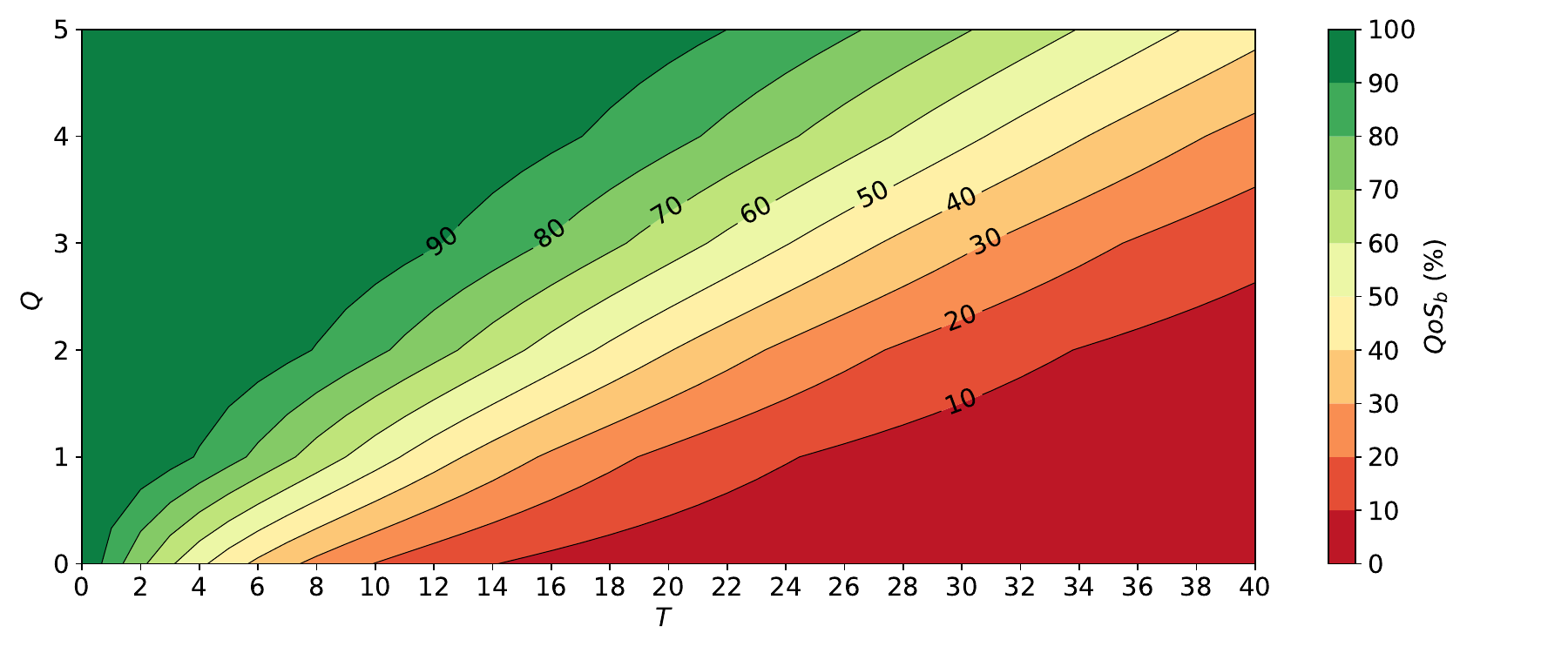}
        \caption{$p_{b}=0.15$}
        \label{fig:QoS_b_0.15p_b}
    \end{subfigure}
    
    \caption{\small{ $QoS_b$ (quality-of-service for prosumers due to consumers' bad behaviour) as a function of $Q$ (number of reserved items in the shared pool) and $T$ (number of prosumers) for two cases of $p_b$ (the probability that a consumer behaves badly). }}
    \label{fig:QoS_b}
\end{figure}

Figures \ref{fig:QoS_ns}, \ref{fig:QoS_ns(QoS_fixed)}, \ref{fig:QoS_s}, and \ref{fig:QoS_b} illustrate how the three decision variables, \( M \), \( T \) and \( Q \), of the optimal system design (Section~\ref{sec:opt}), affect  the relevant $QoS_i$.\\

Figure \ref{fig:QoS_ns(N_fixed)} shows that---for a fixed consumer population (\(N=100\))---\({QoS_{ns}}\) increases with the number of items in the shared pool, (\(M\)),  approaching 100\%. Higher request probabilities, (\(p_{ns}\)), shift the curves to the right, since more items are needed to maintain a particular \({QoS_{ns}}\) when demand is greater. In Figure \ref{fig:QoS_ns(p_ns_fixed)}, with \(p_{ns}\) fixed at 0.10, increasing the consumer population, \(N\), has the same effect:  \(M\) must be increased to maintain the same \({QoS_{ns}}\). In all cases, \({QoS_{ns}}\) is convex for values below 50\% and concave above 50\%, with a maximal slope at 50\%. Near 100\%, the slope becomes very small, demonstrating that pushing \({QoS_{ns}}\) beyond 99\% requires a prohibitive increase in \(M\). \\

Figure \ref{fig:QoS_ns(QoS_fixed)} shows the minimum number, $M$, of items in the shared pool required to maintain  \({QoS_{ns}} \geq99\)\% for different consumer populations, \(N\), and request probabilities, \(p_{ns}\). As \(N\) or \(p_{ns}\) increase ({\em i.e.}\ making greater demand, $X_{ns}$, more probable (\ref{eq:bin})), more items, $M$, are required. Under the normal approximation of the binomial cdf \cite{eusea20244}, which assumes a sufficiently large $N$ and a probability $p_{ns}$ not too close to 0 or 1, $M \approx N p_{ns} + z_{0.99} \sqrt{N p_{ns} (1 - p_{ns})}$ leading to the approximately linear trends observed in Figure~\ref{fig:QoS_ns(QoS_fixed)}.\\

Figure \ref{fig:QoS_s} shows \(QoS_s\) as a heat-map ({\em i.e.}\ contour plot) over the number of items in the shared pool, \(M\), and the number of items made available from prosumers, \(T\) (Figure~\ref{fig:Hybrid}), for four different cases. The green regions represent higher \(QoS_s\), while red indicates poorer $QoS_s$. In Figure \ref{fig:Top-Left}, where \(p_s = 0.3\), \(N = 100\), and \(Q = 1\), increasing either \(M\) or \(T\) smoothly transitions from low to high \({QoS_s}\). Figure \ref{fig:Top-Right} shows that increasing \(p_s\) to 0.4 shrinks the high-\(QoS_s\) ({\em i.e.}\ green) region, meaning more items (\(M\) and/or \(T\)) are needed to maintain a high \(QoS_s\). Meanwhile, comparing Figures \ref{fig:Top-Left} and \ref{fig:Bottom-Left}, an increase in \(N\) from 100 to 150 has the same effect on  \(QoS_s\).  Finally, in Figure \ref{fig:Bottom-Right}, the effect of raising  \(Q\) (the number of reserved items) to 8 again has a similar effect, requiring additional total resources (\(M\) and/or \(T\)) to maintain high $QoS_s$. Note that the contour lines are all linear because  \(QoS_s\) (\ref{eq:bin}) is a function of $A_s \equiv M - Q + T$ (Table~\ref{tab:QoS}) and so is constant for constant  \(M + T\). \\

Figure \ref{fig:QoS_b} shows \({QoS_b}\) as a heat-map over \(Q\) and \(T\), using the same colour conventions as in Figure~\ref{fig:QoS_s}.  

Comparing Figure \ref{fig:QoS_b_0.05p_b} and \ref{fig:QoS_b_0.15p_b}, we note that increasing the  probability of bad behaviour, \(p_b\), rotates  the contours, confirming that the $T$ prosumers require increased resources, $Q$,  to maintain a high  \({QoS_b}\).

The contour lines for a given \({QoS_b}\) are pseudo-linear. To understand why this is so, we note that the binomial distribution with parameters \(\bigl(T, p_b\bigr)\) is well approximated by a normal distribution, ${\mathcal N}(\cdot,\cdot )$, with mean \(T p_b\) and variance \( T p_b(1 - p_b)\). Consequently, the relationship between $Q$ and $T$ for achieving a target $QoS_b$ level can be approximated pseudo-linearly as
\begin{eqnarray}
  Q \approx T\,p_b + y_{QoS_b} \sqrt{T\,p_b\bigl(1 - p_b\bigr)}, \label{eq:Normal_Approximation}  
\end{eqnarray}

where $y_{QoS_b}$ is the quantile of the standard normal distribution corresponding to a specified $QoS_b$, {\em i.e.}\ $P(y \leq y_{QoS_b}) = QoS_b$ for $y \sim \mathcal{ N}(0,1)$. \newline

\begin{remark}[Other QoS functions]
\label{rem:QoS}
In respect of the QoS metric proposed in (\ref{eq:QoS_general_bin}) above, we note the following:
\begin{itemize}
    \item[(a)]
In some situations, empirical QoS functions can be {\em measured\/} directly, overcoming fragile assumptions associated with binomial modelling. 
\item[(b)] (\ref{eq:QoS_general_bin}) represents the overall service level for the entire group rather than the experience of each individual user. While most users may have good access to shared items, some might still face difficulties due to chance. For example, a person may try to use an item several times during busy periods and find it unavailable, even if the system, on average, meets its QoS targets. This randomness means that even in an optimized system, some users might feel that they are receiving a lower level of service. To mitigate this problem, the system could include strategies such as giving priority to those who missed out before. There is a close relationship here to notions of group vs.\ individual fairness in the design of AI decision-making systems~\cite{Binns:20}. 
\end{itemize}
\end{remark}

\section{Mathematical optimization}
\label{sec:opt}

To formalize our approach, we now introduce an optimization framework that determines the most efficient allocation of resources in the hybrid supply scheme (Figure~\ref{fig:Hybrid}). We address two distinct problems. The first focusses on minimising system costs, and the second optimally partitions the shared resources, as follows: 
 
\begin{enumerate}
    \item[] \textbf{Problem 1: Minimum Cost System Design} Given $N$, determine $M$, $T$, and $Q$ in order to minimize a specified cost subject to QoS constraints.
    \item[] \textbf{Problem 2: Best Effort Design} For fixed $N$, $M$, and $T$,  partition the  $M$ shared items to achieve a best possible QoS. For example, this objective might encompass the best overall QoS delivered to consumers and prosumers together; or it might seek to find a partition that equalizes the QoS for consumers and prosumers. In addition,  one can solve this optimization in a privacy-preserving manner without user groups sharing sensitive information. This is related to {\em best effort\/} concepts from the networking community \cite{corless2016aimd}.
\end{enumerate}

\subsection{Problem 1: Minimum Cost System Design}
\label{sec:prob1}
Given a community of $N$ consumers (where $N$ is known), we are interested in determining positive integers, $M$, $T$ and $Q$, to minimize a cost metric  of the following kind:
\begin{eqnarray}
C(M,T,Q) = f(M-Q)+g(Q) + h(T). \label{eq:min}
\end{eqnarray}
Here, the strictly positive scalar functions, $f(M-Q)$, $g(Q)$ and $h(T)$, denote the cost of provision of $M-Q$, $Q$ and $T$ items, respectively (Table~\ref{tab:access_summary}), so that the scalar function, $C(M,T,Q) >0$, is the overall cost of the service. For example, $f(M-Q)$ and $g(Q)$ might encode both the purchase and environmental cost of the shared items and their maintenance, and $h(T)$ the cost of accessing the surge supply, $T$.   For now, we assume that these functions are strictly concave, reflecting the intuition that the marginal cost of provision reduces as the number of items increases ({\em i.e.}\ the economy of scale). This optimization problem is also subject to the following constraints, expressed via (\ref{eq:QoS_general_bin}): 

\begin{eqnarray}
QoS_{ns}(M;N,p_{ns}) &\geq& QoS_{ns,d} \label{eq:nonsurge,d} \\
QoS_{s}(M-Q+T;N,p_{s}) &\geq& QoS_{s,d} \label{eq:surge,d} \\
QoS_{b}(Q;T,p_{b}) &\geq& QoS_{b,d} \label{eq:surge_bad,d}\\
N &\geq& M \label{eq:N>M}\\
M &\geq& Q \label{eq:M>Q}\\
T &\geq& Q \label{eq:T>Q} \\
N&\geq& M-Q+T \label{eq:N>M-Q+T}
\end{eqnarray}

The respective QoS lower bounds, $QoS_{i,d}$, in (\ref{eq:nonsurge,d}--\ref{eq:surge_bad,d}), are pre-specified by the designer (see Table~\ref{tab:QoS}). We will use the sequential least squares 

programming (SLSQP) algorithm~\cite{virtanen2020scipy} to solve this integer-based constrained optimization problem, {\em i.e.}\ to minimize the objective (\ref{eq:min}) as a function of  $M$, $T$, and $Q$, subject to constraints  (\ref{eq:nonsurge,d}--\ref{eq:N>M-Q+T}) .

\subsection{Problem 2: Best Effort Design }
\label{sec:prob2}
A second problem arises when $N$, $M$ and $T$ are fixed and one must select an optimal reserve, $Q$ (Figure~\ref{fig:Hybrid}). This situation arises when one must determine the best possible QoS achievable as a function of $Q$. Alternatively,   in a slowly varying non-stationary environment, one might seek to adjust $Q$ to maintain a given level of QoS. In such situations, one may seek to achieve these goals without sharing the locally specified  $QoS_s$ and $QoS_b$ (see Remark~\ref{rem:QoS}). \newline 

With $N$, $M$, and $T$ fixed, and acknowledging that $QoS_{ns}$ (\ref{eq:nonsurge,d}) does not constrain the optimization variable, $Q$, we rewrite the other two QoS functions (\ref{eq:surge,d},\ref{eq:surge_bad,d}) in terms of $Z\equiv M-Q \geq 0$, via (\ref{eq:QoS_general_bin}): 
\begin{eqnarray}
    QoS_s(Z+T;N,p_s) &\equiv& \mathbb{P}[X_s \leq Z+T | N,p_s], \label{eq:QoS_s(z)} \\
    QoS_b(Q; T,p_b) &\equiv& \mathbb{P}[X_b \leq Q | T,p_b].  \label{eq:QoS_(Q)} 
\end{eqnarray}

$Z\equiv M-Q$  is the number of items (out of the shared $M$ items) that are allocated to the population of $N$ consumers (Figure~\ref{fig:Hybrid}). 

We  want to develop distributed algorithms to solve each of the following problems, respectively: \newline

\textbf{Problem 2a:} For fixed $N$, $M$, and $T$, partition the shared $M$ items to
achieve the best overall QoS delivered to consumers and prosumers together:

\begin{eqnarray}   
   \argmax_Q
   \left\{QoS_s(Z+T;N,p_s)+QoS_b(Q; T,p_b)\right\} \label{eq:Maximize}
\end{eqnarray}

\textbf{Problem 2b:} For fixed $N$, $M$, and $T$, partition the shared $M$ items to equalize the QoS for consumers and prosumers:

\begin{eqnarray}
    \textrm{Equalize}_Q \;\;\;QoS_s(Z+T;N,p_s)=QoS_b(Q; T,p_b) \label{eq:Equalize}
\end{eqnarray}

In these Problem~2 settings, the objectives are therefore defined via the QoS functions themselves (in contrast to Problem~1, in which the QoS functions provide constraints (\ref{eq:nonsurge,d}--\ref{eq:surge_bad,d}) for optimization of an external cost (\ref{eq:min})).   As already mentioned,  distributed solutions are appealing for these problems. First, such algorithms are usually robust to failure in a way that centralized solutions are not. In such situations, distributed (self-organizing) optimization algorithms are very useful. Second, distributed strategies offer the possibility of implementing a variety of policies with minimal requirements for information sharing (privacy issues). \newline

Our approach is to use the AIMD (additive-increase multiplicative decrease) algorithm~\cite{corless2016aimd,wirth2019nonhomogeneous} to solve these problems.  AIMD is a lightweight algorithm (in terms of information exchange between agents) that is used extensively in communication technology as a basis for the TCP/IP protocol~\cite{Schlote2013On}. In the context of Figure~\ref{fig:Hybrid}, consider two {\em agents}, one representing the population, ${\mathcal N}$, of primary users and one representing the population, ${\mathcal T}$, of prosumers. The agents evolve autonomously and respond only to a single-bit feedback signal that indicates when a resource has been allocated. More specifically, agents probe slowly to acquire resource (the additive increase phase), and respond rapidly by releasing resource (the multiplicative decrease phase), once the resource has been allocated. Each agent repeats until a steady-state average allocation has been achieved (either deterministically or stochastically, depending on the implementation). In our hybrid supply scheme (Figure~\ref{fig:Hybrid}), agents probe for the share of the queue of length $M$. Once the equality, $Z+Q=M$, is achieved,  a feedback signal is broadcast to both agents.   In \cite{wirth2019nonhomogeneous}, the authors illustrate how the AIMD algorithm may be modified to solve optimization problems.

\begin{remark}[Convergence of AIMD as an optimization algorithm]
\label{rem:AIMD} Convergence of the AIMD algorithm to solve optimization problems is guaranteed under certain assumptions. The first assumption concerns the nature of the optimal solution. In particular, a necessary assumption is that the optimal point be strictly positive and on the interior of the feasible set. A second assumption concerns the nature of the QoS functions,  $QoS_i$ (\ref{eq:QoS_general_bin}), specifically that these be strictly concave (or convex) \cite{corless2016aimd}. The second of these assumptions is not satisfied by the binomial cdfs, as may be seen in Figure~\ref{fig:QoS_ns}. Nevertheless, (i) measured QoS functions {\em are\/} often concave, reflecting the {\em law of diminishing returns} that typically prevails in these systems; and (ii)  binomial functions can be bounded by a strictly concave function. Interesting questions for future work are to quantify the error under such an approximation;  and to prove AIMD convergence in the case of increasing utility (i.e.\ QoS) functions.

\end{remark}

\begin{algorithm}[t!]
\caption{{\small AIMD algorithm run by consumer and prosumer populations to maximize the overall QoS (Problem 2a)}}
\begin{algorithmic}[1]
\Statex \textbf{Initialization:} The consumer and prosumer populations set their initial states to arbitrary values, $Z(1)$ and $Q(1)$ respectively, and they broadcast parameter, $\Gamma_o>0$, to both agents; 
\Statex $k=0$, $l=0$\\
\textbf{Input:} $M \in {\mathbb N}$, $\alpha > 0$ , $\beta\in (0,1)$  

\For{$l\leftarrow l+1$}
    \If{$Z(l) + Q(l) < M$}
        \State $Z(l+1) = Z(l) + \alpha$
        \State $Q(l+1) = Q(l) + \alpha$
    \Else 
            \State $k\leftarrow k+1$
            \State Evaluate $\bar{Z}(k)$ (\ref{eq:zav}) and $\bar{Q}(k)$ (\ref{eq:qav})
            \State Evaluate $QoS_s'(\bar{Z}(k))$ and $QoS_b'(\bar{Q}(k))$ \ref{eq:binom_pmf}
            \vspace{0.5cm}
            \State $Z(l+1) = \begin{cases}
            \beta \times Z(l) & \text{with probability } \lambda_{o,c}(\bar{Z}(k)) = \Gamma_o\frac{1}{\bar{Z}(k) QoS_s'(\bar{Z}(k))} \\
            Z(l)+\alpha  & \text{with probability } 1 - \lambda_{o,c}(\bar{Z}(k))
            \end{cases}$
            \vspace{0.5cm}
            \State $Q(l+1) = \begin{cases}
            \beta \times Q(l) & \text{with probability } \lambda_{o,p}(\bar{Q}(k)) = \Gamma_o\frac{1}{\bar{Q}(k)QoS_b'(\bar{Q}(k))} \\
            Q(l)+\alpha  & \text{with probability } 1 - \lambda_{o,p}(\bar{Q}(k))
            \end{cases}$
    \EndIf
\EndFor
\end{algorithmic}
\label{alg:1}
\end{algorithm}

\begin{algorithm}[t]
\caption{{\small AIMD algorithm run by consumer and prosumer populations to equalize respective  QoSs (Problem 2b)}}
\begin{algorithmic}[1]
\Statex \textbf{Initialization:} The consumer and prosumer populations set their initial states to arbitrary values, $Z(1)$ and $Q(1)$ respectively, and they broadcast parameter, $\Gamma_e>0$, to both agents; 
\Statex $k=0$, $l=0$\\
\textbf{Input:} $M \in {\mathbb N}$, $\alpha > 0$ , $\beta\in (0,1)$  

\For{$l\leftarrow l+1$}
    \If{$Z(l) + Q(l) < M$}
        \State $Z(l+1) = Z(l) + \alpha$
        \State $Q(l+1) = Q(l) + \alpha$
    \Else     
        \State $k\leftarrow k+1$
        \State Evaluate $\bar{Z}(k)$ (\ref{eq:zav}) and $\bar{Q}(k)$ (\ref{eq:qav})
        \vspace{0.5cm}
        \State $Z(l+1) = \begin{cases}
            \beta \times Z(l) & \text{with probability } \lambda_{e,c}(\bar{Z}(k)) = \Gamma_e\frac{QoS_s(\bar{Z}(k))}{\bar{Z}(k)} \\
            Z(l)+\alpha  & \text{ with probability } 1 - \lambda_{e,c}(\bar{Z}(k))
        \end{cases}$
        \vspace{0.5cm}
        \State $Q(l+1) = \begin{cases}
            \beta \times Q(l) & \text{with probability } \lambda_{e,p}(\bar{Q}(k)) = \Gamma_e\frac{QoS_b(\bar{Q}(k))}{\bar{Q}(k)} \\
            Q(l)+\alpha  & \text{ with probability } 1 - \lambda_{e,p}(\bar{Q}(k))
        \end{cases}$
    \EndIf
   
\EndFor
\end{algorithmic}
\label{alg:2}
\end{algorithm}

Algorithms \ref{alg:1} and \ref{alg:2} show how agents acting for populations $\mathcal N$ and $\mathcal T$ (Figure~\ref{fig:Hybrid}) should follow the policy for the best overall QoS (Problem~2a) or equalized QoS (Problem~2b), respectively.

In these algorithms, $k$ detects and counts cases in which $Z(l)+Q(l)=M$, {\em i.e.}\  the sum of allocations is equal to the total amount of resource available (being the total number of available items, $M$, in the shared pool (Figure~\ref{fig:Hybrid})). In Problem~2a (Algorithm~1), the probability of reducing each population's share (by a factor of $1-\beta$) is given by $\lambda_{o,c}$ for consumers and $\lambda_{o,p}$ for prosumers; while, in Problem~2b (Algorithm~2), these probabilities are  $\lambda_{e,c}$ and $\lambda_{e,p}$, respectively. These probabilities depend, in turn, on the respective average state values. Specifically, if $k\in\{1,2,\ldots\}$ counts the number of cases---labelled $l_j$, $j\in\{1,\ldots,k\}$---in which the constraint, $Z(l_j)+Q(l_j)=M$, is met (we call these capacity events), then the average state values  over these $k\geq 1$ capacity events are, respectively,  
\begin{eqnarray}
\bar{Z}(k)&=&\frac{1}{k}\sum_{j=1}^{k}Z(l_j) = \frac{k-1}{k}\bar{Z}(k-1) + \frac{1}{k}Z(l_k), \label{eq:zav}\\
\bar{Q}(k)&=&\frac{1}{k}\sum_{j=1}^{k}Q(l_j)= \frac{k-1}{k}\bar{Q}(k-1) + \frac{1}{k}Q(l_k). \label{eq:qav}
\end{eqnarray}

 In Algorithm \ref{alg:1}, \( QoS_s'(\cdot) \) and \( QoS_b'(\cdot) \) denote the derivatives of the binomial cdfs, being, therefore, their probability mass functions (pmfs); i.e.

\begin{equation}
QoS'_{i}(m) = \binom{n_i}{m} p_i^{m}(1 - p_i)^{n_i - m}, \;\; m\in \{0,\ldots, n_i\}.
\label{eq:binom_pmf}
\end{equation}

\begin{remark}[Algorithms~\ref{alg:1} and \ref{alg:2}]
Although it appears that the \texttt{for-loop} in Algorithms~\ref{alg:1} and \ref{alg:2} is infinite, this is intentional, reflecting that the agents---consumers and prosumers---continuously follow their assigned policies as long as they use the hybrid supply scheme. Note also that $Z(l)$, $Q(l)$, $\bar{Z}(k)$, and $\bar{Q}(k)$ may not necessarily be integers. Therefore, in our implementation, we employed a continuous approximation for equations~\ref{eq:QoS_s(z)}, \ref{eq:QoS_(Q)}, and \ref{eq:binom_pmf}, the details of which are omitted here for simplicity.
\end{remark}

\section{Application of the hybrid supply scheme to two use cases }
\label{sec:res}
We now instantiate our general hybrid supply sharing economy scheme (Figure~\ref{fig:Hybrid}) in two use cases related to mobility infrastructure sharing. In Section~\ref{sec:caseweight}, we examine a shared mobility scheme that reduces average vehicle weight, deducing the optimal designs specified both by Problem~1 (Section~\ref{sec:prob1}) and Problem~2 (Section~\ref{sec:prob2}). Then, in Section~\ref{sec:casecost}, we propose a shared charging point scheme that reduces cost for a community of EV owners. In this use case, we focus on the optimal design specified by Problem~1 only.  

\subsection{A shared mobility scheme to reduce average vehicle weight}
\label{sec:caseweight}
A major issue in the vehicle industry is the trend toward heavier vehicles \cite{islam2022weight}. This is driven somewhat by safety concerns \cite{monfort2019trends} but primarily by worldwide zero-emission mandates that are encouraging the use of electric vehicles. Electric vehicles are typically heavier than their internal combustion engine (ICE) counterparts due to the ever-increasing consumer demand for range and the associated weight of lithium-ion batteries\cite{Aphale2020Li-ion}. This increased weight is associated with increased non-tailpipe emissions such as those associated with tire wear, brake dust and road dust, and with increased safety risks posed to vulnerable road users \cite{Costagliola2024Non-Exhaust}.\newline   

Our first idea is to explore the cost of a scheme where owners of small electric vehicles automatically have access to a shared pool of larger electric vehicles. Our motivation for this use-case is to not only reduce non-tailpipe emissions but also to reduce demand for DC chargers (associated with large batteries). Since, in almost all counties, on average, longer trips are less frequent than shorter ones, members of this community should access the pool of larger vehicles infrequently \cite{Rohr2013Modeling, Llorca2018Estimation}. More specifically, consider a community of small EV owners who need to have access to large EVs for travelling. In this case:
\begin{itemize}
    \item \textbf{\(N\):} Population of small EV owners.
    \item \textbf{\(M\):} Number of shared large EVs in the pool for $N$ EV owners.
    \item \textbf{\(T\):} Prosumers who own large EV or ICEV and agree to share their cars for $N$ 
    consumers during surge demand.
    \item \textbf{\(Q\):} Reserve of large EVs from the $M$ shared cars as a contingency for the $T$ prosumers.
\end{itemize}

A surge demand scenario can be caused by an exogenous factor such as good weather. In this case, the $M$  shared large EVs are not enough to meet this surge demand. To illustrate our design, we

find $M$, $T$, and $Q$ with a minimum financial cost to satisfy the predetermined QoS ({\em i.e.}\ problem 1). Then, we find optimum $Q$ for the problem 2.

To do that, we need to first determine the Cost and QoS functions.

\paragraph{Cost model:} In 2025, an MG4 EV \footnote{\href{https://www.mg.co.uk/}{mg.co.uk}} costs around £27,000 in the UK and has a range of up to 323 miles, with annual servicing costing £72 (See Table \ref{tab:cost_model}). A typical new car will retain about 40\% of its initial price after 3 years \footnote{\href{https://www.theaa.com/car-buying/depreciation}{theaa.com/car-buying/depreciation}}, meaning it will have depreciated by approximately 60\%. We therefore assume that every 3 years, the fleet will be renewed with the price of $0.6\times £2,7000 = £16,200$ per EV or $£16,200/3=£5,400$ per EV per year. We estimate annual operating costs by including an insurance cost of $£654$ per year \footnote{\href{https://electriccarguide.co.uk/is-it-difficult-to-get-insurance-for-electric-cars/}{electriccarguide.co.uk}} and a parking cost of $£372$ per year \footnote{\href{https://www.richmond.gov.uk/services/parking/parking_permits/car_club_parking_permit/car_club_apply}{richmond.gov.uk/services/parking/}}. This figure only includes fixed infrastructure costs associated with vehicles and does not include any human capital costs (which can easily be incorporated). \\

For modelling prosumer costs, we use daily rental costs to estimate the price paid to prosumers on a monthly basis. Renting a high-range EV \footnote{\href{https://www.enterprisecarclub.co.uk/gb/en/home.html}{enterprisecarclub.co.uk}} starts at £94.79, and we double this on the expectation that prosumers will only ever have to make their cars available twice per month (on average). Thus, we can calculate the financial cost function as the annual cost of operating the scheme: 

\begin{eqnarray}
    C_L(M,T)= 6,500M+12\times200 T
    \label{eq:Car_Cost}
\end{eqnarray}

In equation \ref{eq:Car_Cost}, $C_L(M,T)$ is the annual cost of operating the scheme as a function of $M$ and $T$. The index $L$ shows that this function is linear. However, a more realistic cost function is concave; as we buy more cars, we get a higher discount of up to 25\% maximum \footnote{\href{https://www.privatefleet.com.au/buying-new-car/fleet-discounts/}{privatefleet.com.au/buying-new-car/fleet-discounts/}}. Based on the volume discounts \footnote{\href{https://www.ewaldfleetsolutions.com/blog/multiple-vehicle-discounts/}{ewaldfleetsolutions.com/multiple-vehicle-discounts/},\href{https://www.iracing.com/volume-discounts/}{iracing.com/volume-discounts/}} in Table \ref{tab:Discounts}, we have two cost functions: one is the real cost function (equation \ref{eq:real_cost}) to calculate the real cost, and the second one is the approximated cost function (equation \ref{eq:Aprox_cost}), which is differentiable and is used in the optimization problem.

\begin{eqnarray}
    C_R(M,T)= 6,500 (1-D_R(M))M+12 \times200 T \label{eq:real_cost}\\
    C_A(M,T)=6,500 (1-D_A(M))M+12 \times 200T \label{eq:Aprox_cost}
\end{eqnarray}

In the equations \ref{eq:real_cost} and \ref{eq:Aprox_cost}, $C_R(M,T)$ and $C_A(M,T)$ are annual real and approximated cost functions, respectively. $D_R(M) $ is a piecewise function representing Table \ref{tab:Discounts} for applying discount effects into the real cost function. $D_A(M)$ is a concave differentiable approximation of $D_R(M)$, which makes $C_A(M,T)$ a concave cost function. We consider $D_A(M)$ an exponential decay as

\begin{eqnarray}
    D_A(M)=A(1-e^{-BM}), \label{eq:Approx_Discount}
\end{eqnarray}

where $A$ and $B$ are tuned to produce the best fit curve to the discounts in Table \ref{tab:Discounts}.

\begin{figure}[t]
    \centering
    
    \begin{subfigure}{0.49\textwidth} 
        \centering
        \includegraphics[width=1\linewidth]{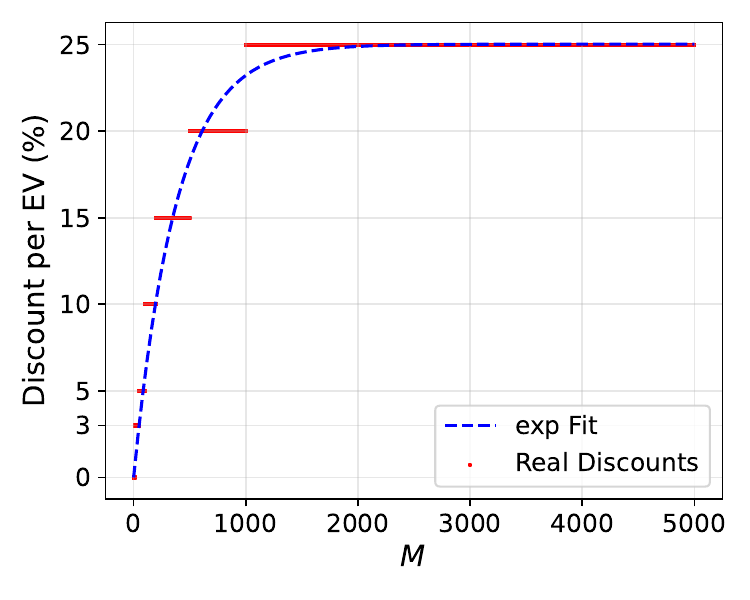} 
        \caption{Discount Functions}
        \label{fig:Aprox_Discount}
    \end{subfigure}
    \hfill
    \begin{subfigure}{0.49\textwidth} 
        \centering
        \includegraphics[width=1\linewidth]{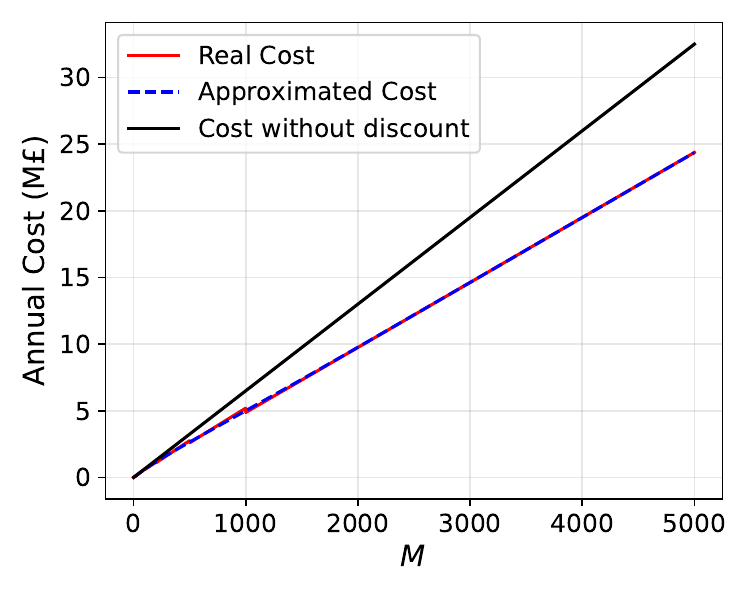} 
        \caption{Cost Functions}
        \label{fig:Three_Different_Cost}
    \end{subfigure}
    
    \caption{{\small (a) shows real discounts and the approximated discounts as exponential fit. (b) shows three cost functions: without discount, which is linear (equation \ref{eq:Car_Cost}); real cost function, which is not convex nor concave (equation \ref{eq:real_cost}); and approximated cost function, which is concave (equation \ref{eq:Aprox_cost}). For these plots, population of prosumers, $T$, is considered 0.}}
    \label{fig:}
\end{figure}

Figure \ref{fig:Aprox_Discount} shows a concave differentiable fit to the real discounts data based on the exponential decay function \ref{eq:Approx_Discount}. Figure \ref{fig:Three_Different_Cost} compares three cost functions: linear \ref{eq:Car_Cost}, real \ref{eq:real_cost}, and approximated \ref{eq:Aprox_cost}. As it can be seen, the approximated one is well matched to the real one, and due to its concavity feature, it has the same slope as the linear one for small values of $M$ but deviates and decreases for large $M$.

\begin{table}[t]
    \centering
    \scriptsize
    \caption{Cost Model for MG4 EV}
    \begin{tabular}{ | l | l | r | }
        \hline
        \textbf{Category} & \textbf{Details} & \textbf{Cost (£)} \\
        \hline
        Vehicle Cost (2025) & MG4 EV & 27,000 \\
        \hline
        Servicing & annual servicing & 72 \\
        \hline
        Depreciation & 60\% of 27,000 over 3 years & £16,200 \\
        \hline

        Annual Fleet Renewal Cost & Per EV per year & 5,400 \\
        \hline
        Insurance & Annual insurance cost per EV & 654 \\
        \hline
        Parking & Annual parking cost per EV & 372 \\
        \hline
        \textbf{Total Annual Costs per EV} & Insurance + Parking + Fleet Renewal+Service & \textbf{$\approx 6,500$} \\
        \hline
    \end{tabular}
    \label{tab:cost_model}
\end{table}

\begin{table}[t]
    \centering
    \scriptsize
    \caption{Discount Structure Based on Quantity}
    \begin{tabular}{|c|c|}
        \hline
        \textbf{Quantity Range} & \textbf{Discount} \\
        \hline
        1 -- 9   & 0\%  \\
        \hline
        10 -- 49  & 3\%  \\
        \hline
        50 -- 99  & 5\%  \\
        \hline
        100 -- 199 & 10\% \\
        \hline
        200 -- 499 & 15\% \\
        \hline
        500 -- 999 & 20\% \\
        \hline
        $>1000$   & 25\% \\
        \hline
    \end{tabular}
    \label{tab:Discounts}
\end{table}

\begin{figure}[t]
    \centering
    \includegraphics[width=1\linewidth]{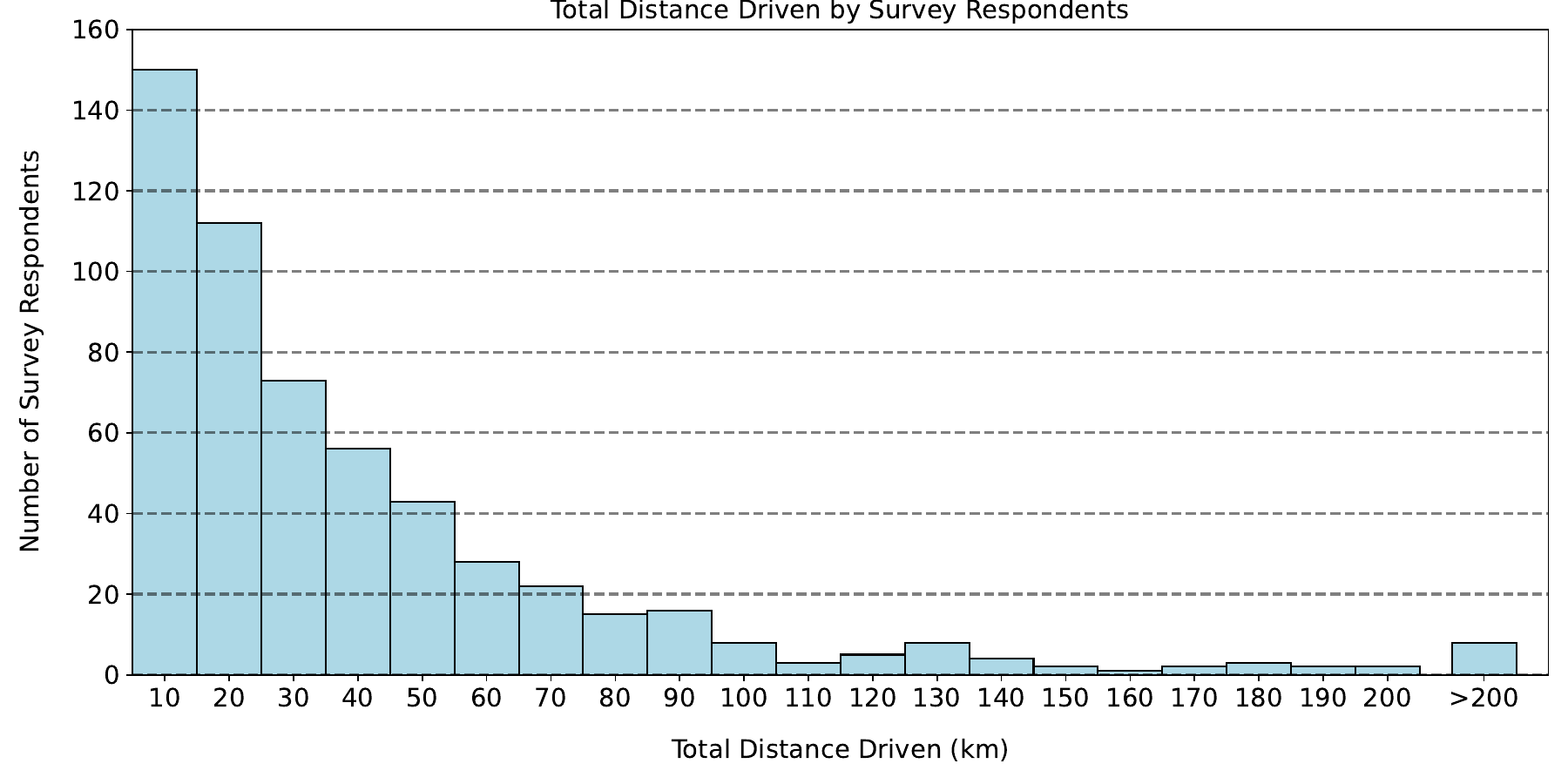}
    \caption{Number of survey respondents reporting about their
travel for the 24 h ‘Monday’ period versus total distances they
drove over that period. \cite{21}}
    \label{fig:Range_Histogram}
\end{figure}

\paragraph{QoS functions:} Based on the data collected from \cite{21}, 

which showed that ~8.86\%–11\% of drivers required trips exceeding 100 km, we estimate the probability for the request for a large EV in non-surge periods is $p_{ns}=0.1$ (see Figure \ref{fig:Range_Histogram}). During surge demand scenarios ({\em e.g.}\ holidays or favourable weather), the probability of high-range EV requests rises to approximately 30\% ($p_s=0.3$).

This estimate aligns with probabilistic EV charging demand models, which report a 25–35\% increase during peak periods due to correlated user activity \cite{ostermann2024probabilistic}, and infrastructure utilization trends in the UK \footnote{\href{https://evpowered.co.uk/feature/electric-vehicle-charging-trends-in-the-uk-2024-and-beyond/}{evpowered.co.uk/feature/electric-vehicle-charging-trends-in-the-uk-2024-and-beyond/}}. The probability of bad behaviour by consumers, such as late returns, can be controlled through penalty fees. We assume that it is set to $p_b=1\%$ by the suppliers.\\

\subsubsection{Results for Problem 1 (Minimum Cost Design): } We simulate a number of scenarios to illustrate the efficacy of our scheme. Specifically, we select $N=\{1000,5000,10000,50000\}$. In each case, we select the $QoS$ constraints to be more than $98\%$ and $99\%$, respectively. 

\subparagraph{Case (i) $N=1000$ :} When $N=1000$ and when the $QoS$ threshold is $98\%$, we find that the optimum values of $M^*=120$,$T^*=216$, and $Q^*=6$, leading to a cost of £$1.22M$ per year. This equates to an additional overall annual cost per consumer of £$1220$. For a higher $QoS$ threshold, such as $99\%$, the optimum values are $M^*=123$,$T^*=217$, and $Q^*=6$, leading to a cost of £$1.24M$ per year or £$1220$ per year per consumer, which are slightly more than $QoS=98\%$.

\subparagraph{Case (ii) $N=5000$ :} When $N=5000$ and when the $QoS$ threshold is $98\%$, we find that the optimum values of $M^*=544$,$T^*=1040$, and $Q^*=17$, leading to a cost of £$5.32$ per year. This equates to an additional overall annual cost per consumer of £$1065$. For a higher $QoS$ threshold, such as $99\%$, the optimum values are $M^*=550$,$T^*=1045$, and $Q^*=19$, leading to a cost of £$5.37M$ per year or £$1074$ per year per consumer, which are slightly more than $QoS=98\%$.

\subparagraph{Case (iii) $N=10000$ :} When $N=10000$ and when the $QoS$ threshold is $98\%$, we find that the optimum values of $M^*=1062$,$T^*=2062$, and $Q^*=30$, leading to a cost of £$10.13$ per year. This equates to an additional overall annual cost per consumer of £$1013$. For a higher $QoS$ threshold, such as $99\%$, the optimum values are $M^*=1070$,$T^*=2069$, and $Q^*=32$, leading to a cost of £$10.18M$ per year or £$1018$ per year per consumer, which is slightly more than $QoS=98\%$.

\subparagraph{Case (iv) $N=50000$ :} When $N=50000$ and when the $QoS$ threshold is $98\%$, we find that the optimum values of $M^*=5138$,$T^*=10196$, and $Q^*=123$, leading to a cost of £$4952$ per year. This equates to an additional overall annual cost per consumer of £$990$. For a higher $QoS$ threshold, such as $99\%$, the optimum values are $M^*=5157$,$T^*=10208$, and $Q^*=126$, leading to a cost of £$49.64M$ per year or £$993$ per year per consumer, which are slightly more than $QoS=98\%$.

Table \ref{tab:_Car_Min_Cost} shows the summary of the results for different scenarios. Increasing $N$ or $QoS$ threshold both lead to an increase in total cost, but by the rise in $N$, cost per consumers decreases due to economy of scale. \newline

\begin{table}[t]
    \centering
    \scriptsize
    \caption{Results for Minimum Cost Design}
    \begin{tabular}{|c|c|c|c|c|c|c|}
        \hline
        \textbf{N} & \textbf{QoS (\%)} & \textbf{M*} & \textbf{T*} & \textbf{Q*} & \textbf{Annual Cost(M£)} & \textbf{Cost per Consumer (£)} \\
        \hline
        \multirow{2}{*}{1000} & 98 & 120 & 216 & 6 & 1.22 & 1220 \\
        \cline{2-7}
        & 99 & 123 & 217 & 6 & 1.24 & 1240 \\
        \hline
        \multirow{2}{*}{5000} & 98 & 544 & 1040 & 17 & 5.32 & 1065 \\
        \cline{2-7}
        & 99 & 550 & 1045 & 19 & 5.37 & 1074 \\
        \hline
        \multirow{2}{*}{10000} & 98 & 1062 & 2062 & 30 & 10.13 & 1013 \\
        \cline{2-7}
        & 99 & 1070 & 2069 & 32 & 10.18 & 1018 \\
        \hline
        \multirow{2}{*}{50000} & 98 & 5138 & 10196 & 123 & 49.52 & 990 \\
        \cline{2-7}
        & 99 & 5157 & 10208 & 126 & 49.64 & 993 \\
        \hline
    \end{tabular}
    \label{tab:_Car_Min_Cost}
\end{table}

\begin{figure}[t!]
    \centering
    
    \begin{subfigure}{0.49\textwidth} 
        \centering
        \includegraphics[width=0.92\linewidth]{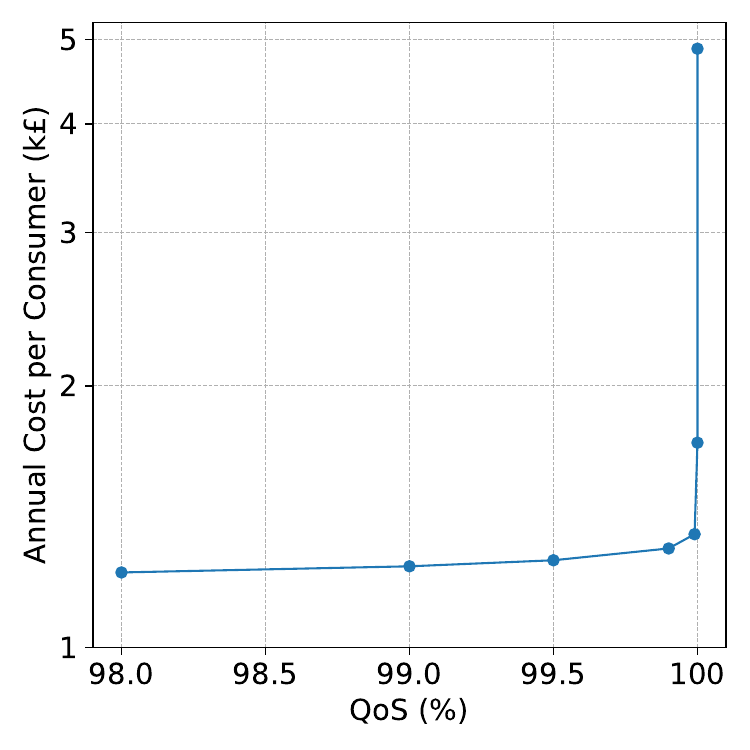} 
        \caption{Cost vs QoS for $N$ = 1000}
        \label{fig:Car_Cost_vs_Qos_N_1000}
    \end{subfigure}
    \begin{subfigure}{0.49\textwidth} 
        \centering
        \includegraphics[width=1.1\linewidth]{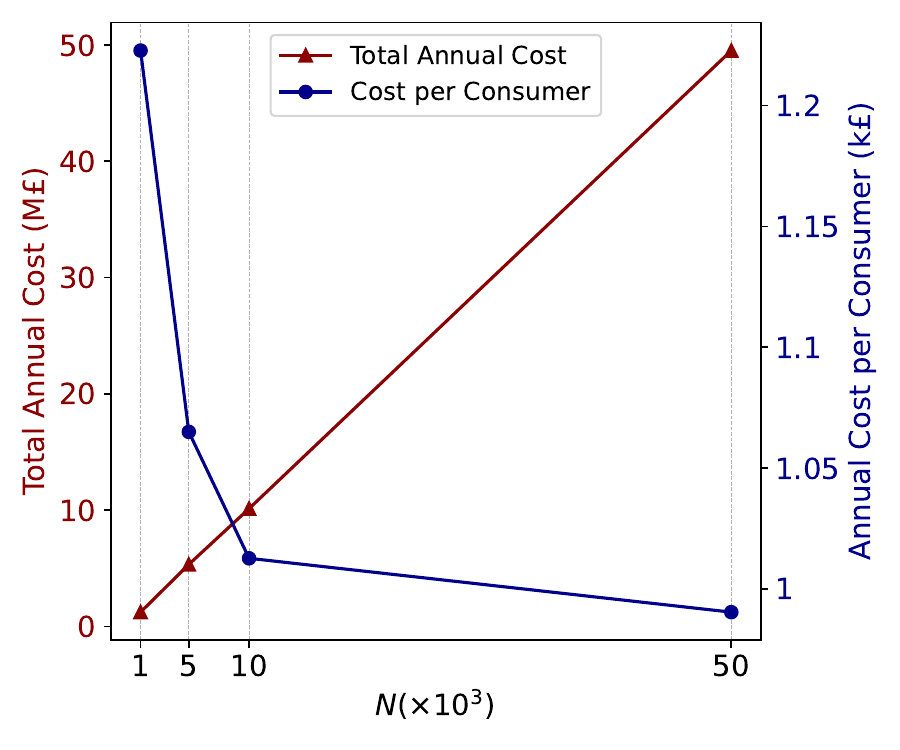} 
        \caption{Cost vs $N$ for QoS = 98\%}
        \label{fig:Cost_vs_N}
    \end{subfigure}
    
    \caption{{\small Annual Cost of the scheme for different values of N (consumers) and QoS (quality of service)}. (a) is the cost-QoS trade-off curve showing the exponential cost of achieving near-perfect QoS, with 100\% QoS being significantly more expensive than slightly lower values for QoS. Note that the Cost axis scale is logarithmic. (b) is the minimum annual cost per N (number of consumers). The left axis shows the total annual cost and the right axis shows the cost per consumer.}
    \label{fig:Cost_Car}
\end{figure}

Figure \ref{fig:Cost_Car} illustrates the relationship between cost and QoS in Figure \ref{fig:Car_Cost_vs_Qos_N_1000}, and cost and $N$ in Figure \ref{fig:Cost_vs_N}. As can be observed in Figure \ref{fig:Car_Cost_vs_Qos_N_1000}, as the QoS approaches 100\%, costs rise sharply. The figure also illustrates that significant cost savings can be achieved while maintaining near-perfect QoS. This highlights the benefits of probabilistic design, where tolerating QoS failure leads to a significant reduction in cost. \ref{fig:Cost_vs_N} shows that although total annual cost increases with an increase in $N$, cost per consumer decreases due to economy of scale. \newline

\begin{remark}[Cost Function]
    Although we used the approximated cost function \ref{eq:Aprox_cost} for the optimisation problem, we used the real cost function \ref{eq:real_cost} to calculate minimum cost values in Table \ref{tab:_Car_Min_Cost} and Figure \ref{fig:Cost_charger}.
\end{remark}

\subparagraph{Comparison to pure B2C :} With the same probability of request inputs, a B2C model can be designed either for non-surge demand periods or surge demand situations. If it is designed only for non-surge demand periods, using $p_{ns}=0.1$ achieving $M = 120$ for $N=1000$ and $QoS \geq 98\%$, then for the surge demand situations there are only two options: either reducing the demand by surge pricing, which is unfair, or accepting the very poor QoS close to zero. Both have disadvantages. If the overall sharing scheme is designed for the worst-case surge scenario, using $p_{s} = 0.3$ we obtain that $M =330$ with a target QoS of 98\%. This is clearly more expensive for consumers and also more wasteful.\\

\begin{figure}
    \centering
    \includegraphics[width=0.8\linewidth]{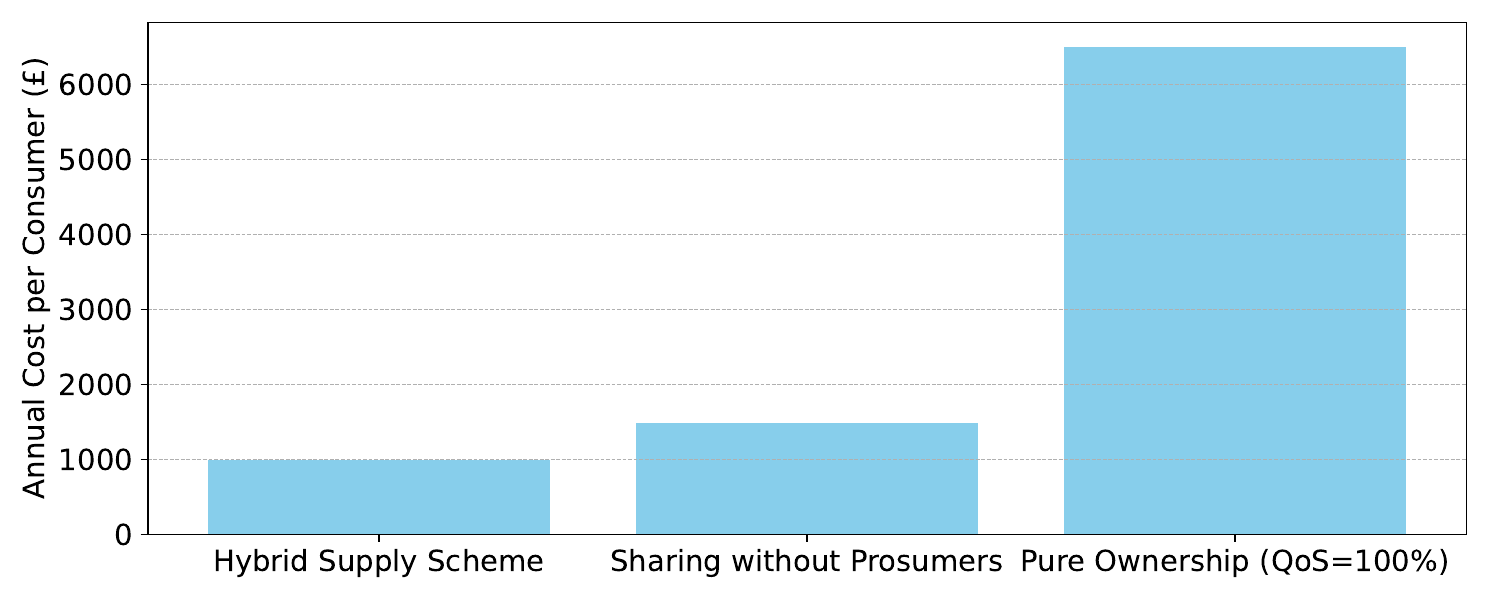}
    \caption{This figure compares the costs of three different approaches for a supply scheme that provides high QoS for customers without surge pricing. This plot is for $N=50,000$}
    \label{fig:Cost_of_Different_Approaches_Car}
\end{figure}

Figure \ref{fig:Cost_of_Different_Approaches_Car} shows a comparison between three approaches; our proposed hybrid supply scheme, sharing without prosumers ({\em i.e.}\ pure B2C), and pure ownership ({\em i.e.}\ $QoS=100\%$). The hybrid supply scheme and pure ownership both provide a high QoS without surge pricing. Here, we assume that the pure B2C is designed in a way that satisfies both criteria of high QoS and no surge demand. As it can be seen, pure B2C is much cheaper than pure ownership, and our proposed hybrid supply scheme is the cheapest one due to incorporating prosumers alongside the shared pool by providing them a reserved part.

\subsubsection{Results for Problem 2 (Best Effort Design): } We explore the best achievable behaviour when $N$, $M$ and $T$ are fixed. For consistency with the above, we select $N=\{1000,5000,10000,50000\}$, and for $M$ and $T$, we select a rounded number close to their optimal values in Table \ref{tab:_Car_Min_Cost}. We then use the AIMD algorithm to find $Q$ such that the overall QoS is maximized (or equalized).\\

Table \ref{tab:AIMD_Car} shows the results for four scenarios. In each scenario, we found the optimum value of $Q$ for both problems 2a maximization (see \ref{eq:Maximize}) and 2b equalization (see \ref{eq:Equalize}). For example, when $N=1000, M=120, T=215$, if we want to achieve the best overall QoS (Maximizing $QoS_s +QoS_b$), the optimal $Q$ is 7, which leads to the average overall QoS 98.66\%. And if we want to achieve equal QoS among both groups of consumers and prosumers, the optimal $Q$ is 5, which leads to the average overall QoS of $97.99\%$. Note that trying to equalize QoS among both groups leads to less average overall QoS, which is in the definition of the problem. It is the decision of the system operator to select which strategy to use. Also, note that with an increase in $N$, a smaller fraction of the shared pool ({\em i.e.}\ $Q^*/M$) needs to be set aside as a reserve part because the scatter binomial cdf of QoS becomes more relaxable.\\

\begin{table}[t]
\centering
\scriptsize
\caption{Results for best effort design}
\begin{tabular}{|l|c|c|c|c|c|c|}
\hline
\multirow{2}{*}{\textbf{Scenario}} & \multirow{2}{*}{\textbf{Problem}} & \multirow{2}{*}{\textbf{Q*}} & \multirow{2}{*}{\textbf{Q*/M (\%)}} & \multicolumn{2}{c|}{\textbf{QoS (\%)}} & \multirow{2}{*}{\textbf{Average (\%)}} \\
\cline{5-6}
& & & & \textbf{QoS\textsubscript{S}} & \textbf{QoS\textsubscript{b}} & \\
\hline
\multirow{2}{*}{N=1000, M=120, T=215} & Maximization & 7 & 5.83 & 97.47 & 99.84 & 98.66 \\
\cline{2-7}
 & Equalization & 5 & 4.17 & 98.17 & 97.8 & 97.99 \\
\hline
\multirow{2}{*}{N=5000, M=545, T=1040} & Maximization & 20 & 3.67 & 97.81 & 99.76 & 98.79 \\
\cline{2-7}
 & Equalization & 17 & 3.12 & 98.24 & 98.04 & 98.14 \\
\hline
\multirow{2}{*}{N=10000, M=1060, T=2065} & Maximization & 34 & 3.21 & 97.68 & 99.77 & 98.73 \\
\cline{2-7}
 & Equalization & 30 & 2.83 & 98.12 & 98.07 & 98.09 \\
\hline
\multirow{2}{*}{N=50000, M=5150, T=10200} & Maximization & 133 & 2.58 & 98.3 & 99.87 & 99.09 \\
\cline{2-7}
 & Equalization & 124 & 2.41 & 98.64 & 98.54 & 98.59 \\
\hline
\end{tabular}
\label{tab:AIMD_Car}
\end{table}

\begin{figure}[h!]
    \centering
    
    \begin{subfigure}{0.49\textwidth} 
        \centering
        \includegraphics[width=1\linewidth]{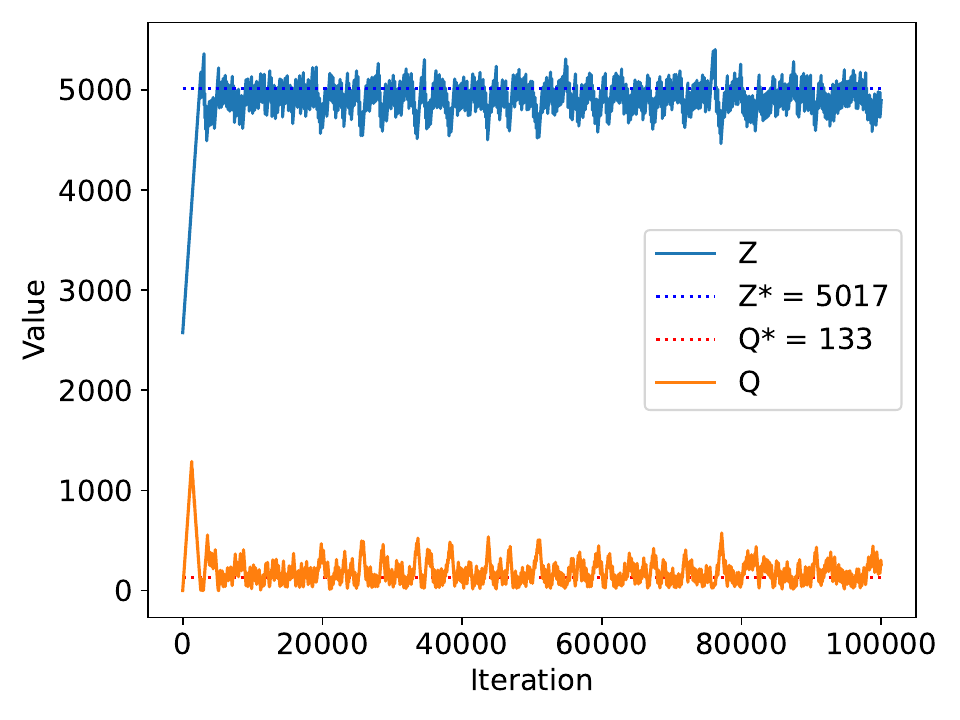} 
        \caption{Values of Z and Q over iterations}
        \label{fig:Z_n_Q_Overall_QoS}
    \end{subfigure}
    \begin{subfigure}{0.49\textwidth} 
        \centering
        \includegraphics[width=1\linewidth]{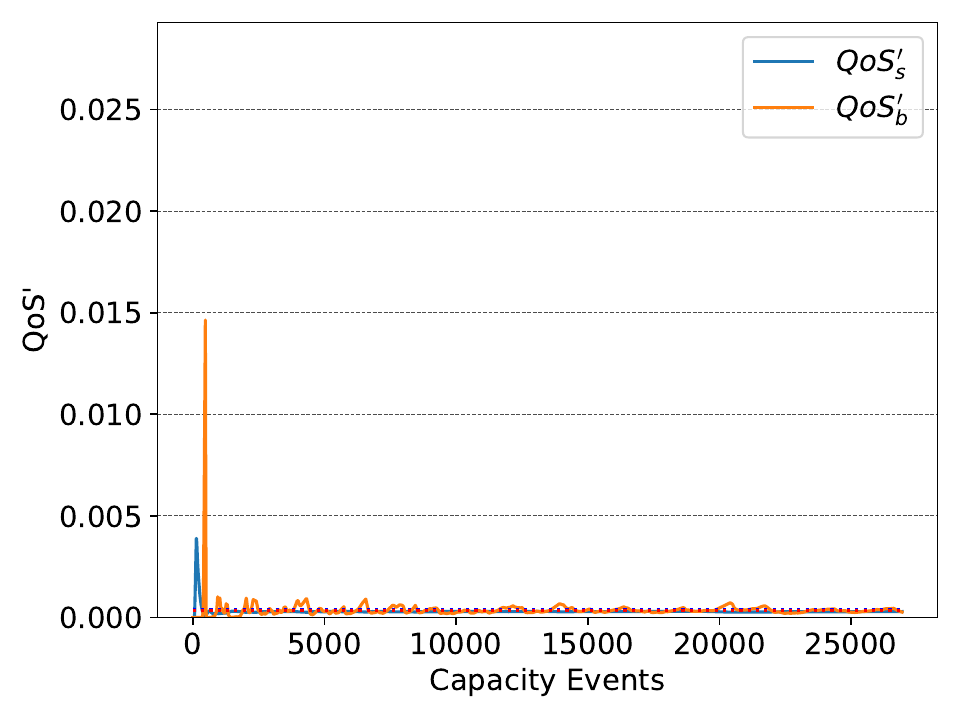} 
        \caption{Evolution of $QoS_s'$ and $QoS_b'$ to consensus}
        \label{fig:d_QoS_Overall_QoS}
    \end{subfigure}
    
     \caption{{\small Convergence of the AIMD algorithm to achieve best overall QoS}}
    \label{fig:Overall_QoS}
\end{figure}

\begin{figure}[h!]
    \centering
    
    \begin{subfigure}{0.49\textwidth} 
        \centering
        \includegraphics[width=1\linewidth]{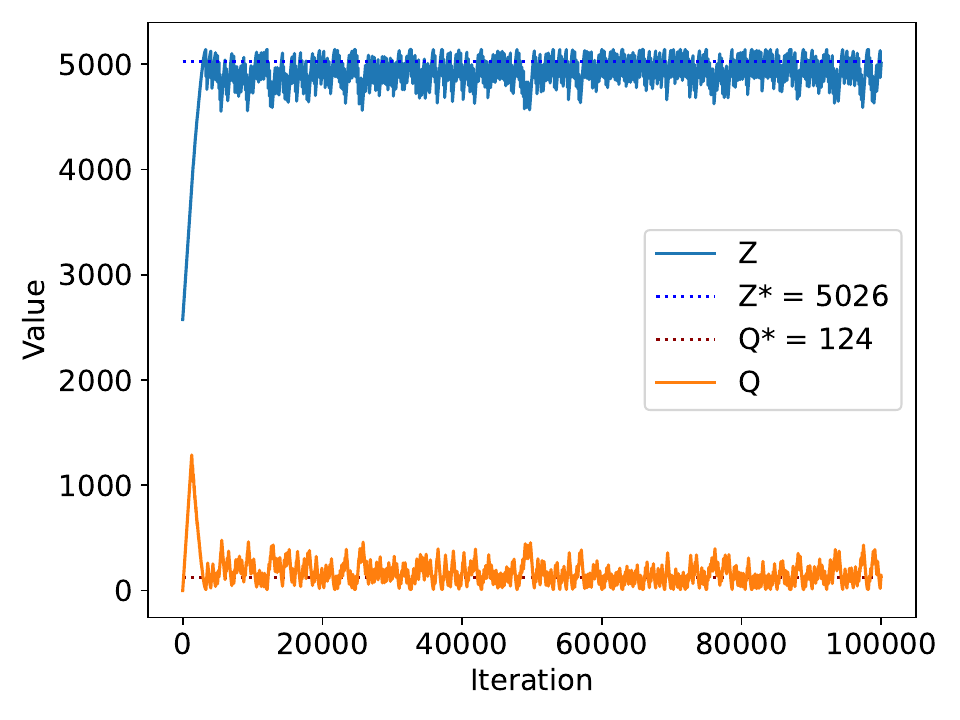} 
        \caption{Values of Z and Q over iterations}
        \label{fig:Z_n_Q_Equal_QoS}
    \end{subfigure}
    \hfill
    \begin{subfigure}{0.49\textwidth} 
        \centering
        \includegraphics[width=1\linewidth]{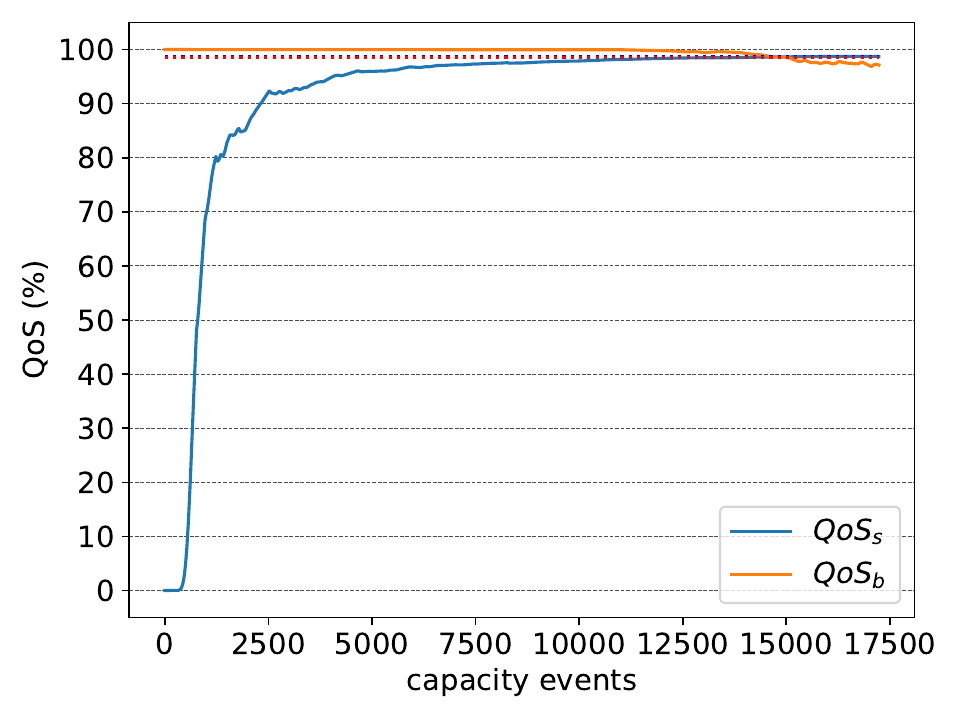} 
        \caption{Evolution of $QoS_s$ and $QoS_b$ to consensus}
        \label{fig:QoS_Equal_QoS}
    \end{subfigure}
    
    \caption{{\small Convergence of the AIMD algorithm to achieve equal QoS among consumers and prosumers}}
    \label{fig:Equal_QoS}
\end{figure}

Figures \ref{fig:Overall_QoS} and \ref{fig:Equal_QoS} show how the AIMD algorithm converges to the optimal value of $Q$ for maximisation and equalization problems, respectively. Figures \ref{fig:Z_n_Q_Overall_QoS} and \ref{fig:Z_n_Q_Equal_QoS} show the fluctuations of $Z$ and $Q$ around their optimal values. Figure \ref{fig:d_QoS_Overall_QoS} shows the evolution of $QoS_s'$ and $QoS_b'$ (equation \ref{eq:binom_pmf}) to consensus, which is the condition for the maximum point of $QoS_s+QoS_b$. Figure \ref{fig:QoS_Equal_QoS} shows the evolution of $QoS_s$ and $QoS_b$ to consensus, which is directly the objective of problem 2b ({\em i.e.}\ equalization \ref{eq:Equalize}).

\subsection{A shared charging point scheme to reduce cost}
\label{sec:casecost}
In our second use case, we develop a system whereby a community of electric vehicle owners shares ownership of a common pool of charging points. In this case:
\begin{itemize}
    \item \textbf{\(N\):} denotes a population of consumers without private charging points.
    \item \textbf{\(M\):} denotes the number of shared charging points for $N$ EV drivers.
    \item \textbf{\(T\):} denotes a population of prosumers who own private charging points and agreed to share their charging points with $N$ EV drivers during surge demand.
    \item \textbf{\(Q\):} is a set reserve of charging points from the $M$ shared charging points that are set aside for prosumers as a contingency situation.
\end{itemize}

For this use case, surges may arise when consumers plan longer trips in a synchronous manner, for example, during periods of good weather at weekends. 

\paragraph{Cost model:} The purchase and installation of a 60 kW DC charger \footnote{\href{https://www.zerovatech.com/}{zerovatech.com}} is around £20,000.

The maintenance cost of a DC charging point is reported \$800 per year \footnote{\href{https://afdc.energy.gov/fuels/electricity-infrastructure-maintenance-and-operation}{afdc.energy.gov/fuels/electricity-infrastructure-maintenance-and-operation}}; {\em i.e.}\ around £54 per month. The estimated price for renting a private charging point for a session through platforms such as Co Charger \footnote{\href{https://co-charger.com/}{co-charger.com}} or Zap-Map \footnote{\href{https://www.zap-map.com/}{zap-map.com}} typically ranges from £5 to £10. Based on these costs, we estimate an average rental income of £20/month per charging point \footnote{\href{https://www.goplugable.com/blog/turn-your-driveway-into-a-passive-income-stream}{goplugable.com/blog/turn-your-driveway-into-a-passive-income-stream}} (two surge periods per month). These costs yield the following linear cost model:

\begin{eqnarray}
    C_L(M,T)=26,480 M + 10\times12\times20 T \label{eq:Charger_Cost}
\end{eqnarray}

Similar to the cost model in the car sharing use case, here we can apply the effect of bulk purchasing using the discounts in table \ref{tab:Discounts_charger} to make the cost function more realistic.

\begin{eqnarray}
    C_R(M,T)=26,480 (1-D_R(M)) M + 10\times12\times20 T \label{eq:Real_Cost_cgarger} \\
    C_A(M,T)=26,480 (1-D_A(M)) M + 10\times12\times20 T \label{eq:Approx_Cost_cgarger}
\end{eqnarray}

Similar to the car sharing use case, $C_R(M,T)$ and $C_A(M,T)$ are annual real and approximated cost functions, respectively. $D_R(M) $ is a piecewise function representing Table \ref{tab:Discounts_charger} for applying discount effects into the real cost function, and $D_A(M)$ is a concave differentiable approximation of $D_R(M)$, which makes $C_A(M,T)$ a concave cost function. $D_A(M)$ is an exponential decay as \ref{eq:Approx_Discount} and $A$ and $B$ are tuned to produce the best fit curve to the discounts in Table \ref{tab:Discounts_charger}.\\

\begin{table}[t]
    \centering
    \scriptsize
    \caption{Discount Structure Based on Quantity for Charging Points}
    \begin{tabular}{|c|c|}
        \hline
        \textbf{Quantity Range} & \textbf{Discount} \\
        \hline
        1 -- 9   & 0\%  \\
        \hline
        10 -- 19  & 5\%  \\
        \hline
        20 -- 49  & 10\%  \\
        \hline
        50 -- 99 & 15\% \\
        \hline
        100 -- 199 & 20\% \\
        \hline
        $>200$   & 25\% \\
        \hline
    \end{tabular}
    \label{tab:Discounts_charger}
\end{table}

\begin{figure}[t!]
    \centering
    
    \begin{subfigure}{0.49\textwidth} 
        \centering
        \includegraphics[width=1\linewidth]{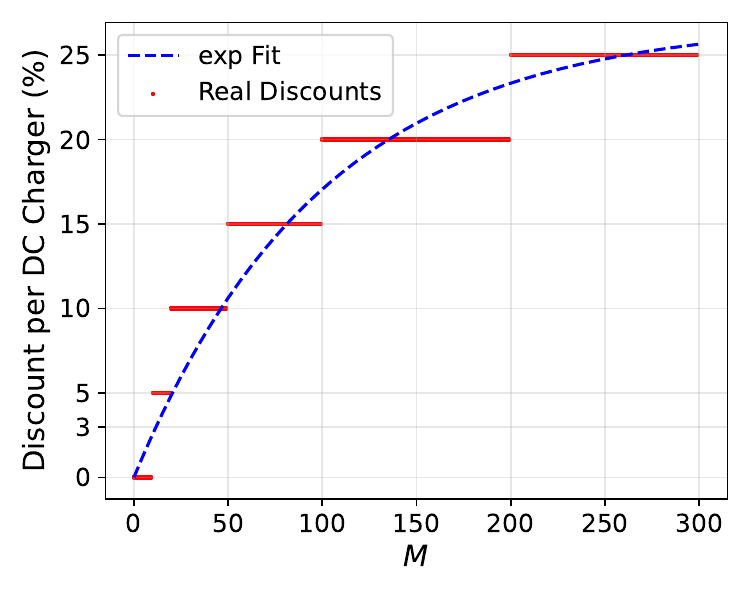} 
        \caption{Discount Functions}
        \label{fig:Aprox_Discount_Charger}
    \end{subfigure}
    \hfill
    \begin{subfigure}{0.49\textwidth} 
        \centering
        \includegraphics[width=1\linewidth]{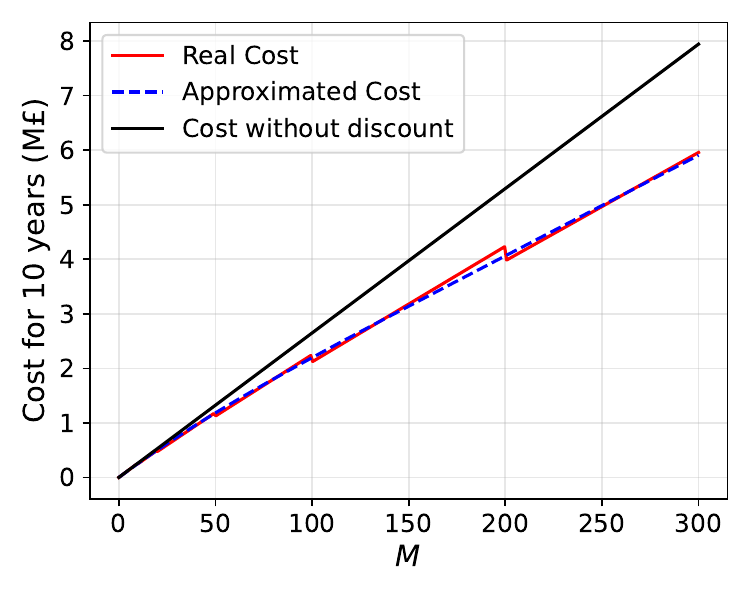} 
        \caption{Cost Functions}
        \label{fig:Three_Different_Cost_Charger}
    \end{subfigure}
    
    \caption{{\small (a) shows real discounts and the approximated discounts as an exponential fit for charging point sharing. (b) shows three cost functions for charging point sharing: without discount, which is linear \ref{eq:Charger_Cost}; real cost function, which is not convex nor concave \ref{eq:Real_Cost_cgarger}; and approximated cost function, which is concave \ref{eq:Approx_Cost_cgarger}. For these plots, population of prosumers, $T$, is considered 0.}}
    \label{fig:cost_charger}
\end{figure}

Figure \ref{fig:Aprox_Discount_Charger} shows the concave differentiable fit to the real discounts data based on an exponential decay function \ref{eq:Approx_Discount}. Figure \ref{fig:Three_Different_Cost_Charger} compares three cost functions: linear \ref{eq:Charger_Cost}, real \ref{eq:Real_Cost_cgarger}, and approximated \ref{eq:Approx_Cost_cgarger}. As it can be seen, the approximated one is well matched to the real one, and due to its concavity feature, it has the same slope as the linear one for small values of $M$ but deviates and decreases for large values of $M$.

\paragraph{QoS functions:} Based on the last data from the UK National Travel Survey (NTS) in 2023, the average driving distance per month is 520 miles per driver. Table \ref{tab:ev_specs} shows the EV distribution among Londoners \footnote{\href{https://www.gov.uk/government/statistics/vehicle-licensing-statistics-2023/vehicle-licensing-statistics-2023}{gov.uk/government/statistics/vehicle-licensing-statistics-2023}}. We calculated the weighted average WLTP (Worldwide Harmonized Light Vehicle Test Procedure) range of popular EV models. This calculation yielded an average range of approximately 284 miles. Given that the average monthly travel distance in the UK is about 520 miles, we estimate that, on average, an EV driver would need to charge about 1.83 times per month. To validate our estimation, we use data provided by the charging company Cirrantic for London in 2024. From this data, the total number of charging sessions recorded by the data provided by Cirrantic for London in 2024 is 2207755. Given that approximately 100 K EV drivers are in London \footnote{\href{https://www.carsloth.com/advice/london-congestion-charge-electric-cars-explained}{carsloth.com/advice/london-congestion-charge-electric-cars-explained}}, we can calculate $2207755/(12\times100k) = 1.84$, which is very close to 1.83 for an average number of charges per month.\\

\begin{table}[t]
    \centering
    \scriptsize
    \caption{Specifications of Top 10 Zero Emission Vehicles Registered in the UK (2023)}
    \label{tab:ev_specs}
    \begin{tabular}{|l|c|c|c|}
        \hline
        \textbf{Model} & \textbf{WLTP Range (miles)} & \textbf{Battery Capacity (kWh)} & \textbf{Number of Registrations} \\
        \hline
        Tesla Model Y       & 315   & 75  & 35,899 \\
        \hline
        MG 4                & 235   & 50  & 21,461 \\
        \hline
        Audi Q4 e-tron      & 265   & 77  & 16,785 \\
        \hline
        Tesla Model 3       & 285   & 65  & 13,547 \\
        \hline
        Polestar 2          & 268   & 78  & 12,540 \\
        \hline
        BMW iX              & 318   & 105 & 11,688 \\
        \hline
        Volkswagen ID.3     & 265   & 58  & 10,265 \\
        \hline
        Kia Niro EV         & 290   & 64  & 10,059 \\
        \hline
        BMW i4              & 318   & 81  & 8,938 \\
        \hline
        Volkswagen ID.4     & 270   & 77  & 8,499 \\
        \hline
    \end{tabular}
\end{table}

\begin{figure}[t]
    \centering
    \includegraphics[width=1\linewidth]{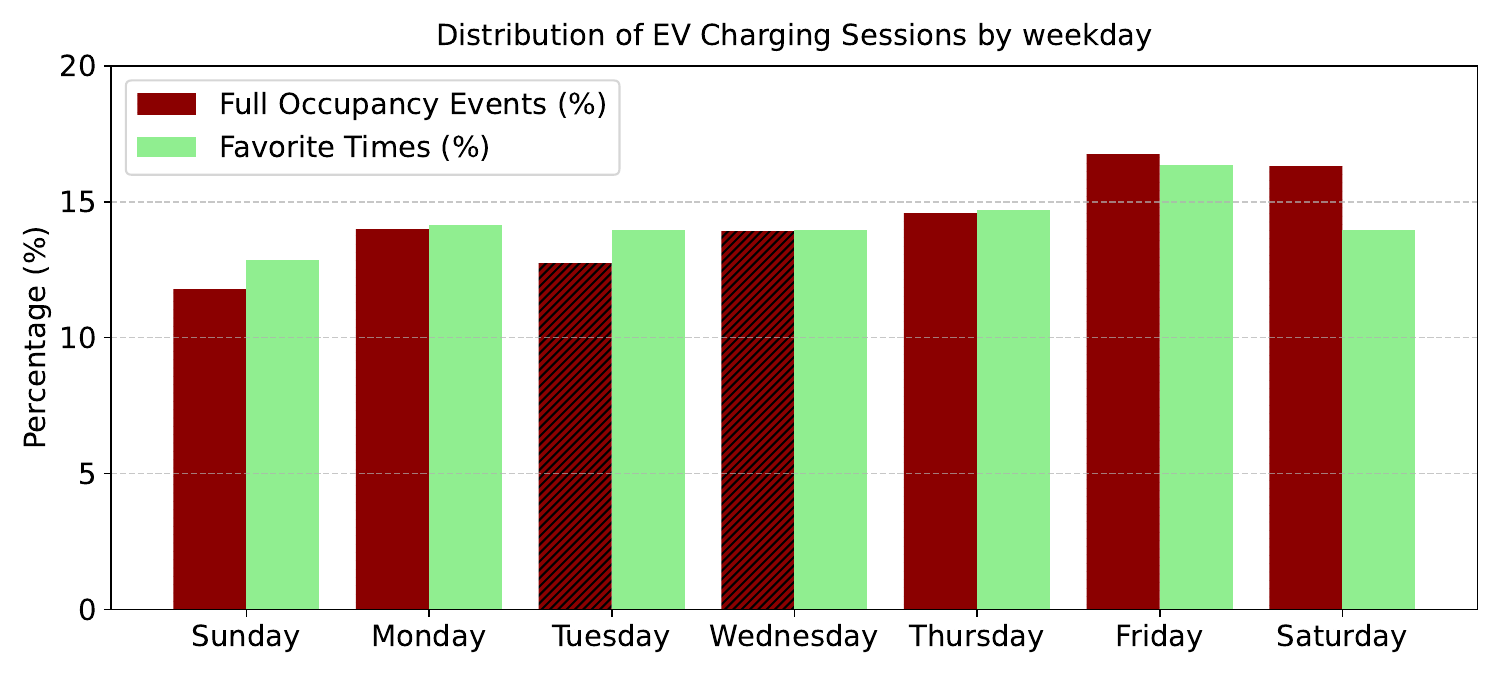}
    \caption{Distribution of EV Charging Sessions by Weekday. The bar chart presents the percentage distribution of EV charging sessions across different days of the week. It highlights full occupancy events (hatched bars) and favourite charging times (green bars). These data have been provided by Cirrantic.}
    \label{fig:weekday}
\end{figure}

\begin{figure}[t]
    \centering
    \includegraphics[width=1\linewidth]{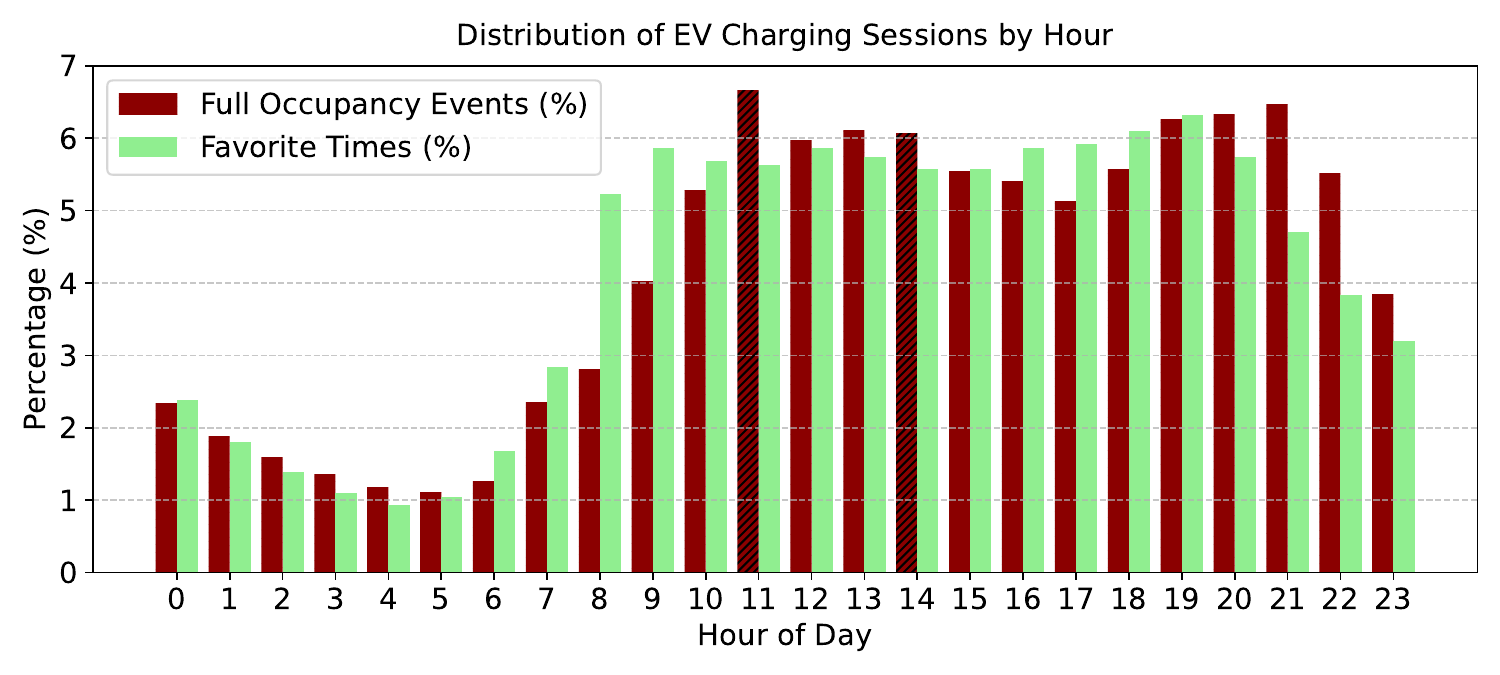}
    \caption{Distribution of EV Charging Sessions by Hour. The bar chart illustrates the percentage distribution of EV charging sessions throughout the day, highlighting full occupancy events (hatched bars) and favourite charging times (green bars). These data have been provided by Cirrantic.}
    \label{fig:Charging-point-Data}
\end{figure}

\begin{figure}
    \centering
    \includegraphics[width=1\linewidth]{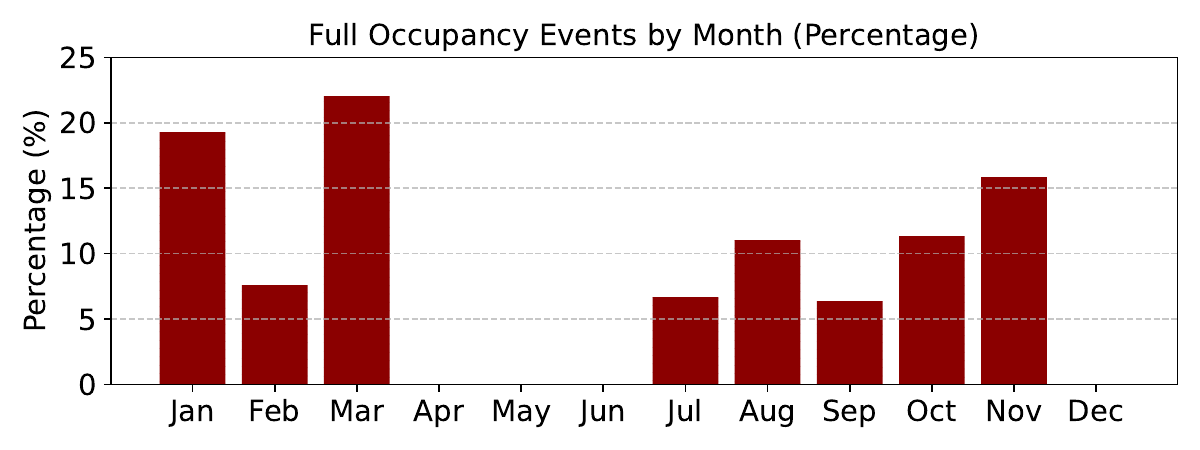}
    \caption{Distribution of Full Occupancy Events by Month. The bar chart illustrates the percentage of full occupancy events across different months of the year. The data shows peak congestion in January, March, and November, while months such as April, May, and June experience little to no full occupancy events. This suggests seasonal variations in EV charging demand. These data have been provided by Cirrantic.}
    \label{fig:full_occupancy_by_month}
\end{figure}

Figure \ref{fig:weekday} illustrates that the utilization of charging points is relatively uniform across weekdays. Figure \ref{fig:Charging-point-Data} depicts which hours of the day are more favoured for charging. As is expected, day-time hours from 8 am to 8 pm experience more charging activity than other periods.  We estimate the per-hour probability of a charging session for $\mathcal{N}$ to be $p_{ns}= 100\times1.83/360 = 0.5\%$. This is based on the fact that there are, on average, 1.83 charges each month, which is approximately 30 days, and $30\times12=360$ hours for DC sessions. To estimate $p_s$, we use dates from Figure \ref{fig:full_occupancy_by_month}, where we observe maximum demand mostly happens in March; namely, so $p_s$ is $1.5\%$. \\

[t]

\subsubsection{Results for Problem 1 (Minimum Cost Design): } The reported probability of bad consumer behaviour $p_b$ is stated as $15\%$ by CIRRANTIC and leads to a very expensive scheme. To handle this large value of $p_b$, we suggest using a deposit-based scheme to encourage consumers to vacate prosumer charge points as contracted to achieve $p_b=1\%$. In the following experiments, we simulate a number of scenarios to illustrate the efficacy of our scheme. Specifically, we select $N=\{1000,5000,10000,50000\}$. In each case, we select the $QoS$ constraints to be more than $98\%$ and $99\%$, respectively. 

\subparagraph{Case (i) $N=1000$ :} When $N=1000$ and when the $QoS$ threshold is $98\%$, we find that the optimum values of $M^*=10$,$T^*=14$, and $Q^*=1$, leading to a total cost of £$0.29$M over 10 years. This equates to an additional overall annual cost per consumer of £$28.56$. For a higher $QoS$ threshold, such as $99\%$, the optimum values are $M^*=11$,$T^*=15$, and $Q^*=1$, leading to a total cost of £$0.31$M over 10 years or £$31.27$ per year per consumer, which is slightly more than $QoS=98\%$.

\subparagraph{Case (ii) $N=5000$ :} When $N=5000$ and when the $QoS$ threshold is $98\%$, we find that the optimum values of $M^*=36$,$T^*=60$, and $Q^*=3$, leading to a total cost of £$1.00$M over 10 years. This equates to an additional overall annual cost per consumer of £$20.04$. For a higher $QoS$ threshold, such as $99\%$, the optimum values are $M^*=37$,$T^*=62$, and $Q^*=3$, leading to a total cost of £$1.03$M over 10 years or £$20.61$ per year per consumer, which is slightly more than $QoS=98\%$.

\subparagraph{Case (iii) $N=10000$ :} When $N=10000$ and when the $QoS$ threshold is $98\%$, we find that the optimum values of $M^*=65$,$T^*=114$, and $Q^*=4$, leading to a total cost of £$1.74$M over 10 years. This equates to an additional overall annual cost per consumer of £$17.37$. For a higher $QoS$ threshold, such as $99\%$, the optimum values are $M^*=67$,$T^*=116$, and $Q^*=4$, leading to a total cost of £$1.79$M over 10 years or £$17.86$ per year per consumer, which is slightly more than $QoS=98\%$.

\subparagraph{Case (iv) $N=50000$ :} When $N=50000$ and when the $QoS$ threshold is $98\%$, we find that the optimum values of $M^*=283$,$T^*=534$, and $Q^*=11$, leading to a total cost of £$6.90$M over 10 years. This equates to an additional overall annual cost per consumer of £$13.80$. For a higher $QoS$ threshold, such as $99\%$, the optimum values are $M^*=287$,$T^*=538$, and $Q^*=11$, leading to a total cost of £$6.99$M over 10 years or £$13.98$ per year per consumer, which is slightly more than $QoS=98\%$.\\

Table \ref{tab:cost_analysis_charger} shows the summary of the results for different scenarios. Increasing $N$ or $QoS$ threshold both lead to an increase in total cost, but by the rise in $N$, cost per consumers decreases due to economy of scale. \newline

\begin{table}[t]
    \centering
    \scriptsize
    \caption{Results for Minimum Cost Design}
    \begin{tabular}{|c|c|c|c|c|c|c|}
        \hline
        \textbf{N} & \textbf{QoS (\%)} & \textbf{M*} & \textbf{T*} & \textbf{Q*} & \textbf{Cost(M£) over 10 years} & \textbf{Annual Cost per Consumer (£)} \\
        \hline
        \multirow{2}{*}{1000} & 98 & 10 & 14 & 1 & 0.29 & 28.56  \\
        \cline{2-7}
        & 99 & 11 & 15 & 1 & 0.31 & 31.27 \\
        \hline
        \multirow{2}{*}{5000} & 98 & 36 & 60 & 3 & 1.00 & 20.04 \\
        \cline{2-7}
        & 99 & 37 & 62 & 3 & 1.03 & 20.61  \\
        \hline
        \multirow{2}{*}{10000} & 98 & 65 & 114 & 4 & 1.74 & 17.37  \\
        \cline{2-7}
        & 99 & 67 & 116 & 4 &  1.79 & 17.86 \\
        \hline
        \multirow{2}{*}{50000} & 98 & 283 & 534 & 11 &  6.90 & 13.80  \\
        \cline{2-7}
        & 99 & 287 & 538 & 11 & 6.99 & 13.98  \\
        \hline
    \end{tabular}
    \label{tab:cost_analysis_charger}
\end{table}

\begin{figure}[t]
    \centering
    
    \begin{subfigure}{0.49\textwidth} 
        \centering
        \includegraphics[width=0.92\linewidth]{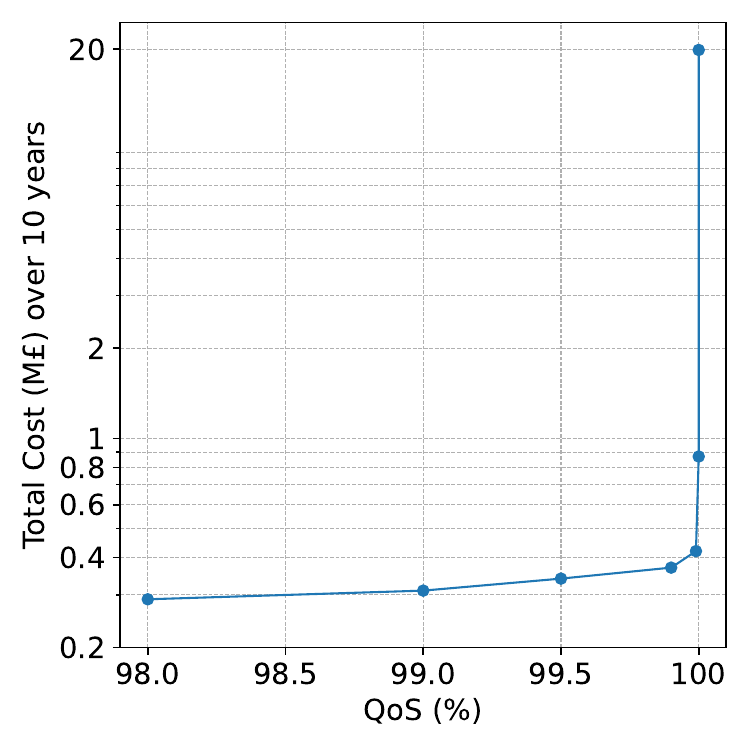} 
        \caption{Cost vs QoS for $N$ = 1000}
        \label{fig:Charger_Cost_vs_Qos_N_1000}
    \end{subfigure}
    \begin{subfigure}{0.49\textwidth} 
        \centering
        \includegraphics[width=1.1\linewidth]{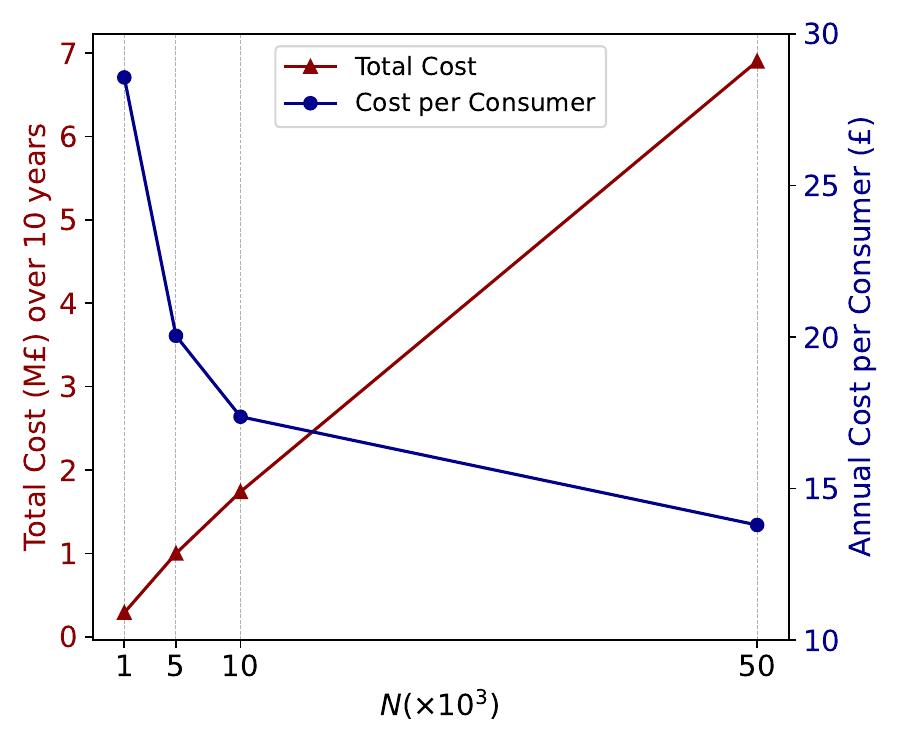} 
        \caption{Cost vs $N$ for QoS = 98\%}
        \label{fig:Cost_vs_N_charger}
    \end{subfigure}
    
    \caption{{\small Cost of the scheme for different values of N (consumers) and QoS (quality of service)}. (a) is the cost-QoS trade-off curve showing the exponential cost of achieving near-perfect QoS, with 100\% QoS being significantly more expensive than slightly lower values for QoS. Note that the Cost axis scale is logarithmic. (b) is the minimum cost vs N. The left axis shows the total cost over 10 years, and the right axis shows the annual cost per consumer.}
    \label{fig:Cost_charger}
\end{figure}

Figure \ref{fig:Cost_charger} illustrates the relationship between cost and QoS in \ref{fig:Charger_Cost_vs_Qos_N_1000} and cost and $N$ in \ref{fig:Cost_vs_N_charger}. As can be observed in \ref{fig:Charger_Cost_vs_Qos_N_1000}, as the QoS approaches 100\%, costs rise sharply. The figure also illustrates that significant cost savings can be achieved while maintaining near-perfect QoS. This highlights the benefits of probabilistic design, where tolerating QoS failure leads to a significant reduction in cost. \ref{fig:Cost_vs_N_charger} shows that although total cost increases with an increase in $N$, cost per consumer decreases due to economy of scale. \newline

\subparagraph{Comparison to pure B2C :} With the same probability of request inputs, a B2C model can be designed either for non-surge demand periods or surge demand situations. If it is designed only for non-surge demand periods, using $p_{ns}=0.5\%$ achieving $M = 10$ for $N=1000$ and $QoS \geq 98\%$, then for the surge demand situations there are only two options: either reducing the demand by surge pricing, which is unfair, or accepting the very poor QoS close to zero. Both have disadvantages. If the overall sharing scheme is designed for the worst case surge scenario, using $p_{s} = 1.5\%$, we obtain that $M =23$ with a target QoS of 98\%. This is clearly more expensive for consumers and also more wasteful.\\

\begin{figure}[t]
    \centering
    \includegraphics[width=0.8\linewidth]{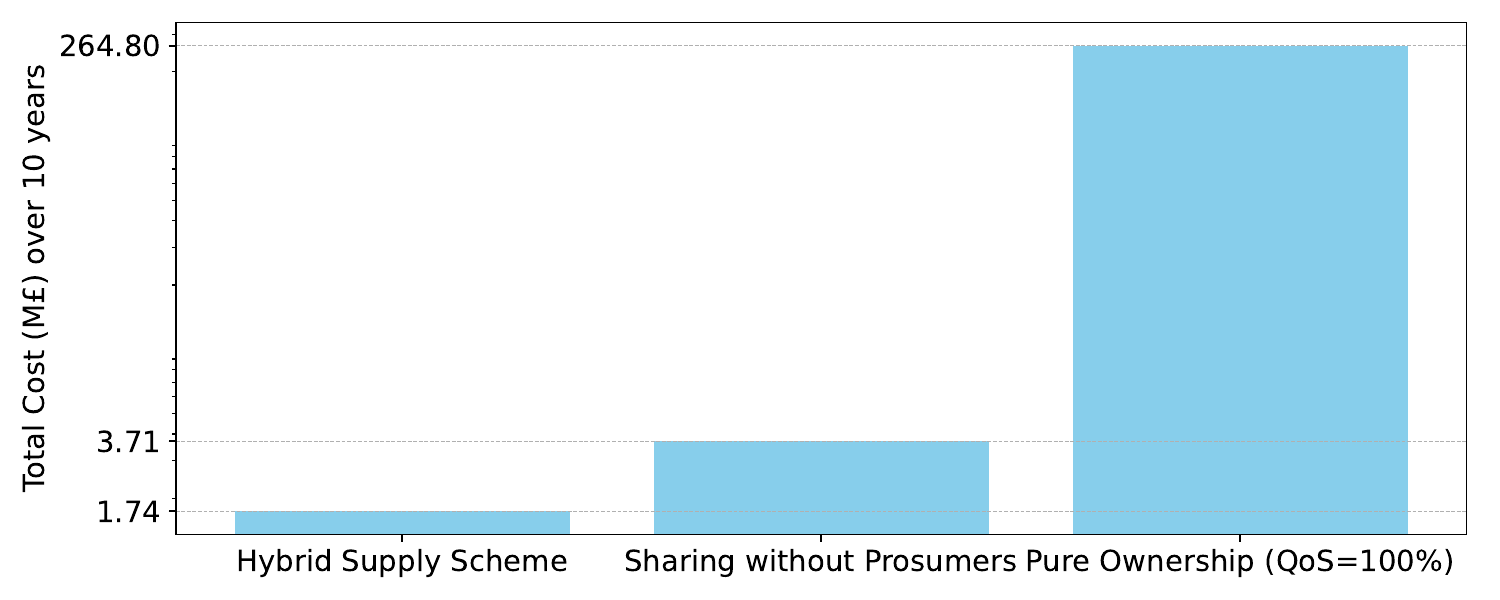}
    \caption{This figure compares the costs of three different approaches for a supply scheme that provides high QoS for consumers without surge pricing. Y axis scale is logarithmic.}
    \label{fig:Cost_of_Different_Approaches_charger}
\end{figure}

Figure \ref{fig:Cost_of_Different_Approaches_charger} shows a comparison between three approaches; our proposed hybrid supply scheme, sharing without prosumers ({\em i.e.}\ pure B2C), and pure ownership ({\em i.e.}\ $QoS=100\%$). The hybrid supply scheme and pure ownership both provide a high QoS without surge pricing. Here, we assume that the pure B2C is designed in a way that satisfies both criteria of high QoS and no surge demand. As it can be seen, pure B2C is much cheaper than pure ownership, and our proposed hybrid supply scheme is the cheapest one due to incorporating prosumers alongside the shared pool by providing them a reserved part.

\section{Conclusion}
\label{sec:con}

The sharing economy systems frequently need to accommodate surge demand, which traditional approaches address either by restricting access, resulting in low quality-of-service (QoS) or through surge pricing, a method that disadvantages less affluent consumers. In this paper, we propose a novel hybrid supply scheme that leverages surge sourcing from resource owners (prosumers) to deliver high QoS to consumers without relying on surge pricing. We incorporate a reserve portion of the shared pool exclusively for prosumers mitigating resource access issues through unpredictable rental return events. The coupling between the shared pool and its reserve was achieved by solving two problems: a minimum cost system design and a best effort design with fixed resources. \newline

We apply this framework to two use cases for car sharing and charging point sharing. We demonstrate that our hybrid supply scheme is cheaper than alternative solutions in providing high QoS without surge pricing and that resources are more efficiently allocated. By adopting a probabilistic approach to define QoS, we underscore the advantages of probabilistic design in sharing economy systems, offering a resource-efficient and accessible alternative to conventional methods. This work highlights the potential of surge sourcing to optimize resource use and support the sharing economy’s goals of accessibility and sustainability.

\section*{Acknowledgments}
We acknowledge Cirrantic \footnote{\href{https://cirrantic.com/}{cirrantic.com}} for providing the charge point data.

\section*{Declaration of Interest}
The authors declare no competing financial interests or personal relationships that could have influenced the work reported in this paper.

\section*{Funding}
This work was supported by the UKRI underwrite of the European Union’s Horizon Europe Marie Skłodowska-Curie Actions (MSCA) under grant agreement number 101073508. 

\section*{Data Availability Statement}
Data will be made available on request.

\bibliographystyle{elsarticle-num} 
\bibliography{cas-refs}

\end{document}